\documentclass[fleqn,usenatbib]{mnras}

\usepackage{newtxtext,newtxmath}

\usepackage[T1]{fontenc}

\usepackage[normalem]{ulem} 

\DeclareRobustCommand{\VAN}[3]{#2}
\let\VANthebibliography\thebibliography
\def\thebibliography{\DeclareRobustCommand{\VAN}[3]{##3}\VANthebibliography}

\usepackage{graphicx}	
\usepackage{color}
\usepackage{booktabs}
\usepackage{breakcites}
\usepackage[table]{xcolor}
\usepackage{empheq}

\mathchardef\mhyphen="2D

\newcommand{\di}{\mathrm{d}}
\newcommand{\bfx}{\mathbf{x}}

\newcommand{\bfv}{\mathbf{v}}

\newcommand{\pc}{\,{\rm pc}}
\newcommand{\kpc}{\,{\rm kpc}}
\newcommand{\Myr}{\,{\rm Myr}}

\newcommand{\kms}{\,{\rm km\, s^{-1}}}

\newcommand{\cs}{c_{\rm s}}
\newcommand{\cg}{c_{\rm g}}

\newcommand{\hatetheta}{\hat{\textbf{e}}_\theta}

\newcommand{\pa}{\partial}
\newcommand{\Msun}{\, \rm M_\odot}
\newcommand{\Msunyr}{\, \rm M_\odot yr^{-1}}
\newcommand{\hatex}{\hat{\textbf{e}}_x}
\newcommand{\hatey}{\hat{\textbf{e}}_y}

\newcommand{\Omegap}{\Omega_{\rm p}}
\newcommand{\hatephi}{\hat{\mathrm{\mathbf{e}}}_\phi}

\title[]{Nuclear rings are the inner edge of a gap around the Lindblad Resonance}

\author[Sormani et al.]{
Mattia C.\ Sormani,$^{1}$\thanks{E-mail: mattiacarlo.sormani@gmail.com}
Emanuele Sobacchi,$^{2}$
and Jason L.\ Sanders$^{3}$
\\
$^1$ Department of Physics, University of Surrey, Guildford GU2 7XH, UK \\
$^2$ INAF – Osservatorio Astronomico di Brera, via E. Bianchi 46, I–23807 Merate, Italy \\
$^3$ Department of Physics and Astronomy, University College London, London WC1E 6BT, UK
}

\date{Accepted XXX. Received YYY; in original form ZZZ}
\pubyear{2015}

\begin{document}
\label{firstpage}
\pagerange{\pageref{firstpage}--\pageref{lastpage}}
\maketitle

\begin{abstract}
Gaseous nuclear rings are large-scale coherent structures commonly found at the centres of barred galaxies. We propose that they are an accumulation of gas at the inner edge of an extensive gap that forms around the Inner Lindblad Resonance (ILR). The gap initially opens because the bar potential excites strong trailing waves near the ILR, which remove angular momentum from the gas disc and transport the gas inwards. The gap then widens because the bar potential continuously excites trailing waves at the inner edge of the gap, which remove further angular momentum, moving the edge further inwards until it stops at a distance of several wavelengths from the ILR. The gas accumulating at the inner edge of the gap forms the nuclear ring. The speed at which the gap edge moves and its final distance from the ILR strongly depend on the sound speed, explaining the puzzling dependence of the nuclear ring radius on the sound speed in simulations.
\end{abstract}



\section{Introduction} \label{sec:introduction}

Gaseous nuclear rings are remarkable structures commonly found at the centres of barred galaxies. They have typical radii of $50\mhyphen1000\pc$ \citep{Comeron+2010}, total gas masses of $10^8\mhyphen10^9 \Msun$ \citep{Sheth2005,Querejeta2021}, and star formation rates spanning a wide range $0.1\mhyphen10\Msunyr$ \citep{Mazzuca+2008,Ma2018}. They are among the most intense star-forming regions of disc galaxies and are considered special laboratories to study star formation under extreme conditions \citep{Moon2021,Schinnerer2023}. They are sites where galactic outflows can be launched, with profound impact on the evolution of their host galaxies \citep{Veilleux2020}. They constitute cold gas reservoirs for the fuelling of central supermassive black holes. The Milky Way hosts a nuclear ring with a radius of $R\simeq120\pc$ that is better known as the Central Molecular Zone \citep{MorrisSerabyn1996,Henshaw2022}.

It is well-known that nuclear rings are easy to form in simulations (e.g.\ \citealt{Athan92b,Kim++2012a,Sormani2015a}, and many others). The recipe is simple: let gas flow in a non-axisymmetric rotating barred potential, and a nuclear ring will spontaneously form in the central regions. In the simplest simulations, the gas is assumed to be 2D, isothermal, non-self gravitating, and the barred potential is externally imposed, but a ring can form also if additional physics is included, for example the gas self-gravity, star formation \& stellar feedback, live stellar potentials, or magnetic fields \citep{Fux1999,Armillotta2019,Tress2020}. However, being able to watch the ring forming in simulations does not mean that we understand the underlying physical process by which it forms, which has remained elusive.

Despite the interest from several astrophysical communities, the physical mechanism by which nuclear rings form is not well understood. What sets the radius of the nuclear ring? What is ``special'' about its location? Various theories have been proposed, but we argue that they all fail to explain the formation of the rings. These previous theories are reviewed in Sect.~\ref{sec:previoustheories}.

One of the most puzzling aspects is that the radius of the nuclear ring in isothermal simulations of gas flow in a barred potential depends very strongly on the assumed sound speed \citep[e.g.][]{Englmaier1997,Patsis2000,Kim++2012a,Sormani2015a}. For example, doubling the sound speed from $\cs=5\kms$ to $10\kms$ can change the radius of the ring by a factor of two or more (see for example Figure 2 in \citealt{Sormani2015a}). This is surprising because the sound speed always amounts to just a small fraction of the orbital speed (typically $\sim$5\%). The flow is always strongly supersonic. None of the currently available theories can explain the strong dependence of the ring radius on the sound speed.

In this paper, we develop a framework to understand the formation of nuclear rings. We propose that the rings are in fact the inner edge of an extensive gap that opens around the ILR due to the excitation of waves by a bar potential. These waves remove angular momentum from the gas disc, transporting the gas inwards. The nuclear ring forms due to the accumulation of gas at the inner edge of the gap. 

The paper is structured as follows. In Section~\ref{sec:numericalexperiments} we present some numerical experiments that illustrate the formation of nuclear rings in simulations. In Section~\ref{sec:previous} we review the constraints that we believe any plausible theory for the formation of nuclear rings should satisfy, and we review previous theories. In Section~\ref{sec:lineardisc} we study the excitation of density waves by an external bar potential using linear theory. In Section~\ref{sec:formation} we illustrate our picture of the formation of the rings. In Section~\ref{sec:discussion} we discuss various connections between this paper and previous works, in particular the works of \cite{Goldreich1978,Goldreich1979} that studied the opening of the Cassini gap in Saturn's rings. We sum up in Section~\ref{sec:conclusion}.

\section{Numerical experiments} \label{sec:numericalexperiments}

We first perform some numerical experiments by letting non-self-gravitating isothermal gas flow in an external barred potential. This is useful to establish some key points and parameter dependencies that we will need later.

\subsection{Numerical setup} \label{sec:numerics}

We run a total of six 2D non-self-gravitating isothermal simulations of gas flowing in an external barred gravitational potential. The potential is described in Appendix~\ref{appendix:potential}. Table~\ref{tab:sims} provides a summary of the simulations run. The equations of motion are:
\begin{align} 
& \pa_t \rho + \nabla \cdot (\rho \bfv) = 0 \,, 								 \label{eq:continuity} \\
 & \pa_t \bfv + (\bfv \cdot \nabla) \bfv = - \frac{ \nabla P}{\rho}  - \nabla \Phi \,,	\label{eq:euler}
\end{align}
where $\rho$ is the surface density, $\bfv$ is the gas velocity, $\Phi$ is the external gravitational potential given by Eq.~\eqref{eq:potential} and
\begin{equation} \label{eq:isothermaleos}
    P = \cs^2 \rho\,,
\end{equation}
is the isothermal equation of state, where $\cs={\rm constant}$. We will use values in the range $\cs=1\mhyphen20\kms$.

We solve Eqs.~\eqref{eq:continuity} and \eqref{eq:euler} using the public grid code {\sc PLUTO} \citep{Mignone2007} on a two-dimensional
static polar grid in the region $R \times \theta =  [0.1 \kpc, R_{\rm max}] \times [0, 2\pi]$. The grid is logarithmically spaced in $R$ and uniformly spaced in $\theta$ with $N_R \times 1024$ cells. The resolution along the $R$ direction is approximately $\Delta R=0.00529~R$, i.e. we have a resolution of $\Delta R=0.529\pc$ at the inner boundary of $R=100\pc$. The number of cells in each direction is chosen so that the aspect ratio of the cells is approximately $\Delta \theta (R/\Delta R)\sim 1$. We use the following parameters: {\sc rk2} time-stepping, no dimensional splitting, {\sc hll} Riemann solver and the default flux limiter. We solve the equations in the frame rotating at $\Omegap$ by using the {\sc rotating\_frame = yes} switch. Boundary conditions are outflow both on the inner boundary at $R=0.1\kpc$ and on the outer boundary at $R=R_{\rm max}$.

The initial density distribution is 
\begin{equation}
    \rho_0 = \begin{cases}
         \bar{\rho} \quad \,\,\, \text{if } R \leq R_{\rm disc}\,, \\
        \rho_\epsilon \quad \text{if } R > R_{\rm disc}\,.
    \end{cases}
\end{equation}
Note that, since the equations of motion \eqref{eq:continuity} and \eqref{eq:euler} are invariant under density rescaling, the density units are arbitrary. The quantity $\bar{\rho}$ therefore essentially sets the density units, and without loss of generality we set $\bar{\rho}=1$. The quantity $\rho_\epsilon=10^{-12}\bar{\rho}$ corresponds to the density floor imposed in the simulation to avoid crashing. We introduce the bar gradually to reduce transients \citep[e.g.][]{Athan92b}. We start with gas in equilibrium on circular orbits in the logarithmic axisymmetric potential $\Phi_0$ and then linearly turn on the non-axisymmetric part of the potential $\Phi_1$ during the first $313\Myr$.

\subsection{Disc with initial radius smaller than the ILR} \label{sec:numericalresults1}

\begin{figure*}
	\includegraphics[width=\textwidth]{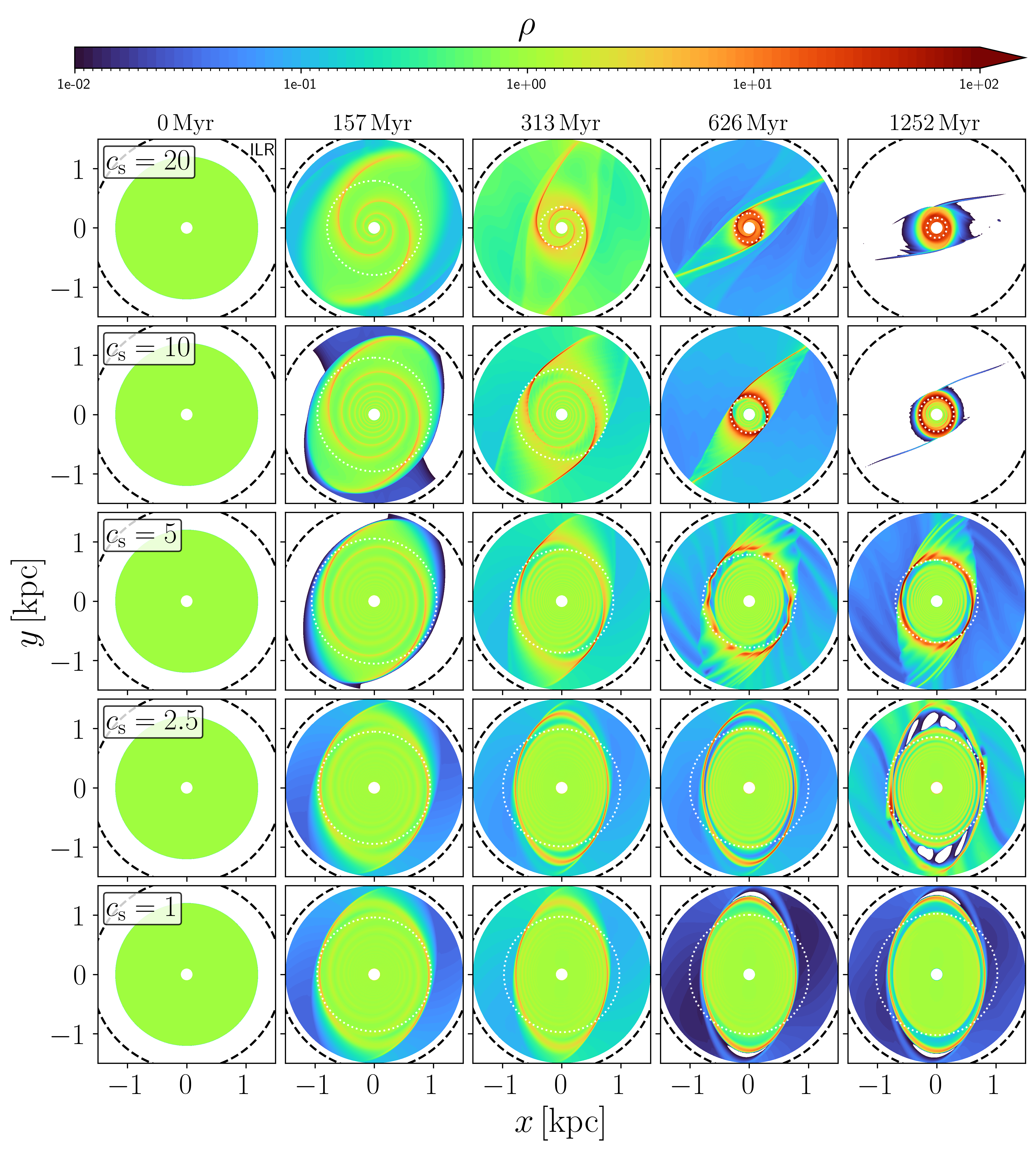}
    \caption{Surface density of simulations 01-05 (see Table~\ref{tab:sims}), illustrating the formation of nuclear rings for various values of the sound speed $\cs$. The only difference between these simulations is the assumed $\cs$. Time increases from left to right. Sound speed decreases from top to bottom (top row is the 05 simulation, bottom row is the 01 simulation). The black dashed circle indicates the ILR. The white dotted circle indicates the instantaneous ring radius according to the definition used in Fig.~\ref{fig:ringsize}. All panels are rotated so that the major axis of the bar potential (i.e. the $\theta=0$ line in Eq.~\ref{eq:potential}) coincides with the $x$ axis. The sense of rotation is clockwise. Trailing density waves excited by the bar potential are visible (see in particular the second and third column from the left). The radius of the ring at the end of the simulation (rightmost column) strongly depends on the sound speed. Regions with densities $\rho<10^{-2}$ are shown white.}
    \label{fig:manyrho}
\end{figure*}

Simulations 01-05 investigate the evolution of a uniform gas disc with an initial radius $R_{\rm disc}=1.2\kpc$ that is smaller than $R_{\rm ILR}=1.61\kpc$ (Appendix~\ref{appendix:potential}). The only difference between these five simulations is the assumed sound speed (Table~\ref{tab:sims}). Figure \ref{fig:manyrho} shows the surface density as a function of time.

As soon as the bar potential is turned on, trailing spiral waves are excited. These waves are clearly visible at $t=157 \Myr$ and $t=313 \Myr$. The movies of the surface density as a function of time show that the waves are first excited at the outer edge of the disc, and propagate inwards. We will confirm later in Sects.~\ref{sec:excitation} and \ref{sec:analyticedge} using linear analysis that sharp edges are indeed regions where strong wave excitation takes place, and therefore play a key role in the formation of the rings. The wiggles that are visible along the spirals in some panels (for example the panel at $t=313\Myr$ and $\cs=10\kms$) are due to the wiggle instability \citep{WadaKoda2004,KimKimKim2014,Sormani+2017a,Mandowara2022}.

\begin{figure}
   	\includegraphics[width=\columnwidth]{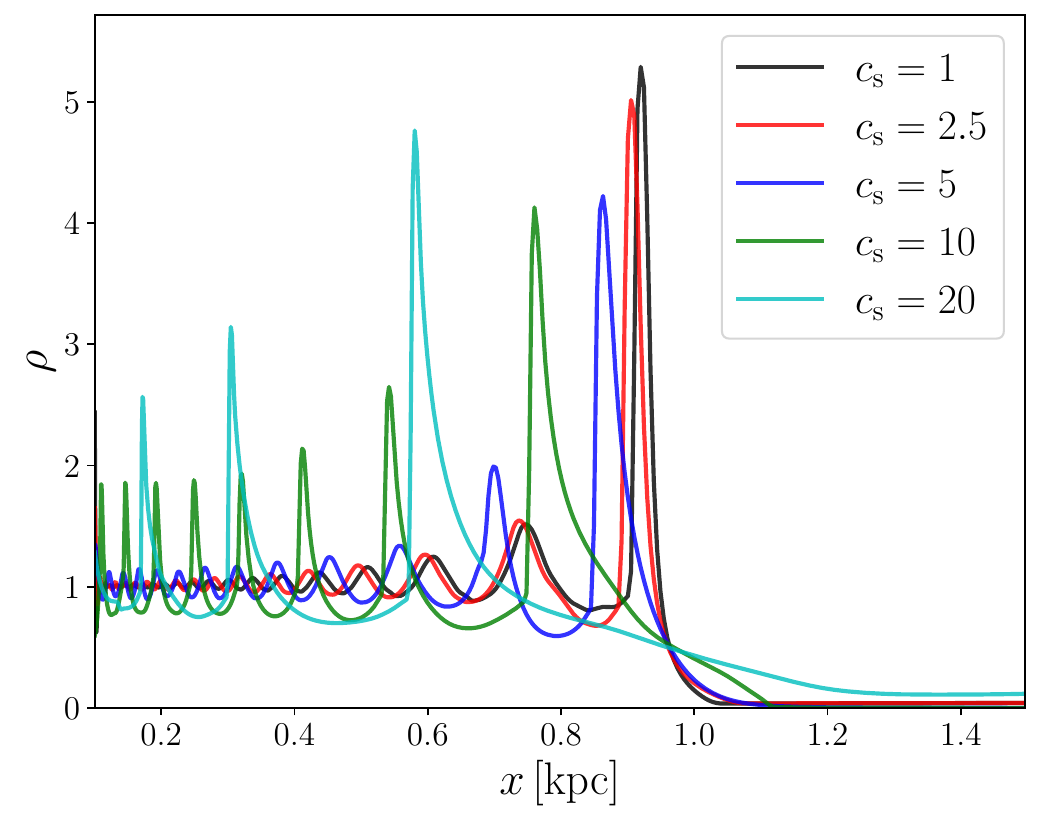}
    \includegraphics[width=\columnwidth]{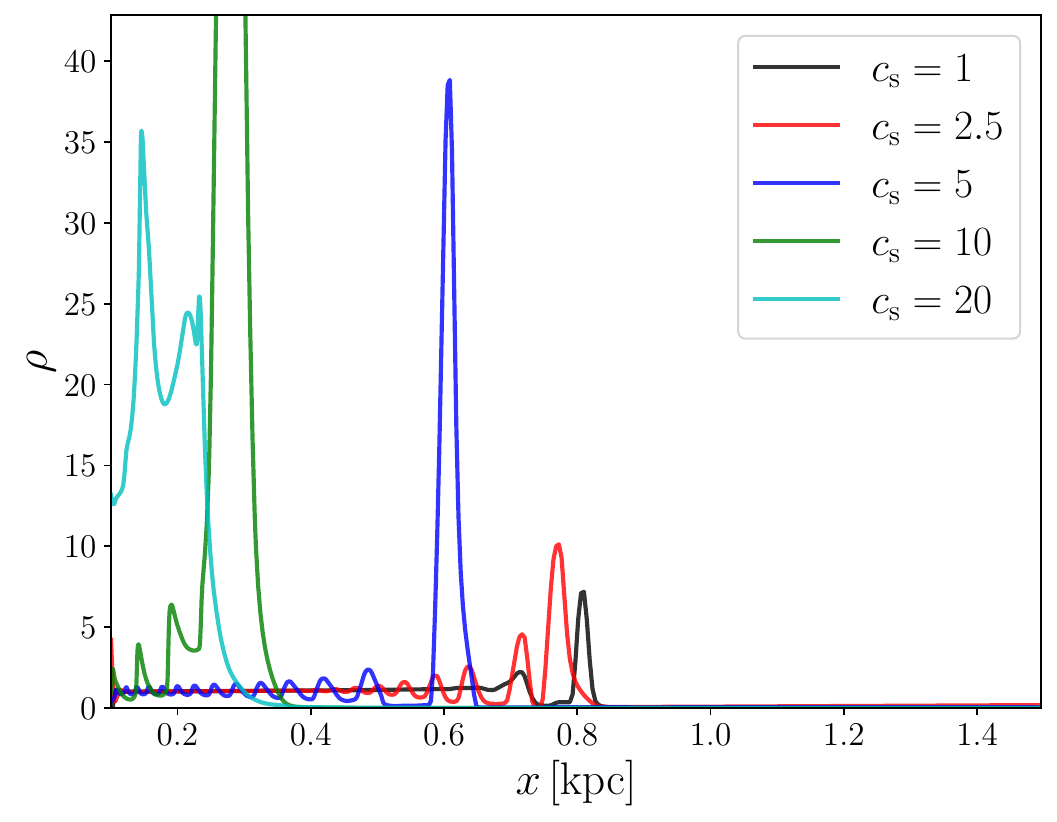}
    \caption{Surface density on the $x$ axis for simulations 01-05 at $t=157\Myr$ (top) and and $t=1252\Myr$ (bottom). In other words, these are horizontal cuts in the second column and fifth column of Fig.~\ref{fig:manyrho}. The oscillations are the density waves excited by the bar potential. The wavelength of the waves increases with increasing sound speed $\cs$. The waves are highly non-linear.}
    \label{fig:xcut}
\end{figure}

Figure \ref{fig:xcut} plots a cut through the $x$ axis of Fig.~\ref{fig:manyrho} at $t=157\Myr$. The radial wavelength increases with increasing sound speed. This will be explained by the dispersion relation derived below (Eq.~\ref{eq:dispersionrelation}). The amplitude of the waves decreases inward, despite the prediction of the linear analysis according to which the amplitude of density waves should increase inward due to geometric effects (see Sect~\ref{sec:lineardisc}). The reason for this behaviour is that the waves in the simulations become quickly non-linear and develop shocks. The shocks cause the waves to dissipate, decreasing their amplitude and depositing their (negative) angular momentum into the gas disc. As we will argue in Sect.~\ref{sec:formation}, this process is what decreases the angular momentum of the gas disc and causes it to shrink.

\begin{figure}
	\includegraphics[width=\columnwidth]{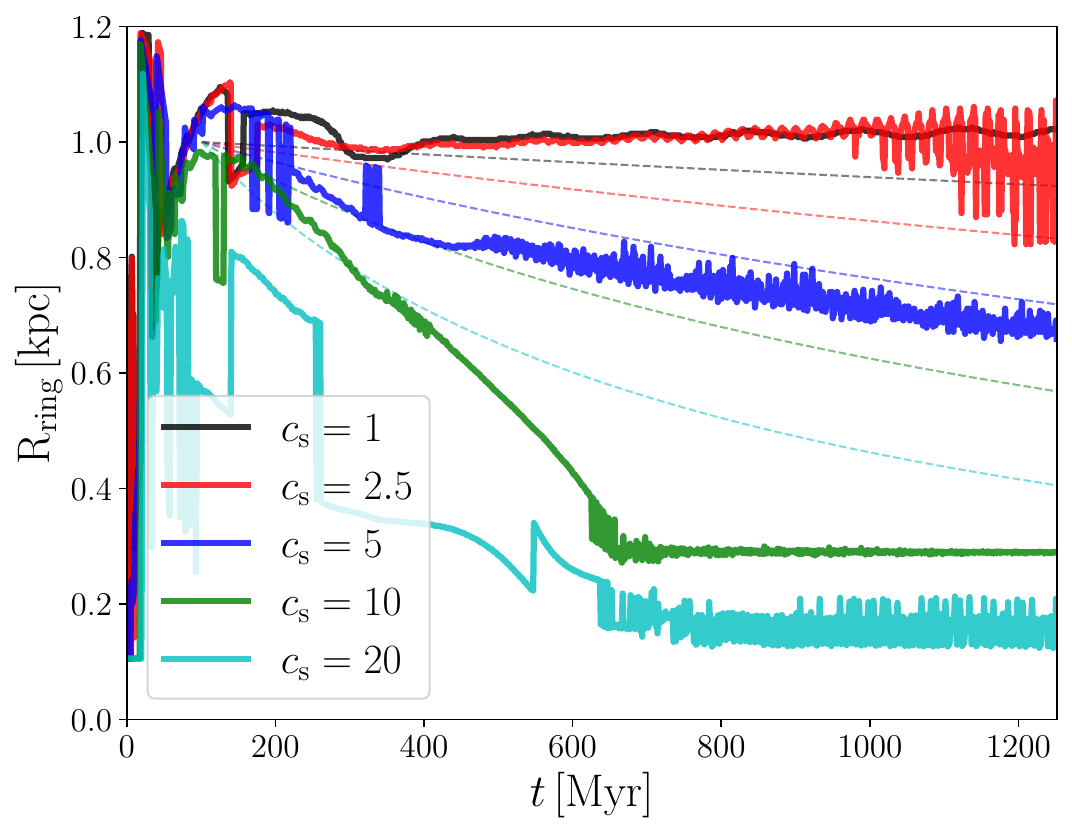}
 	\includegraphics[width=\columnwidth]{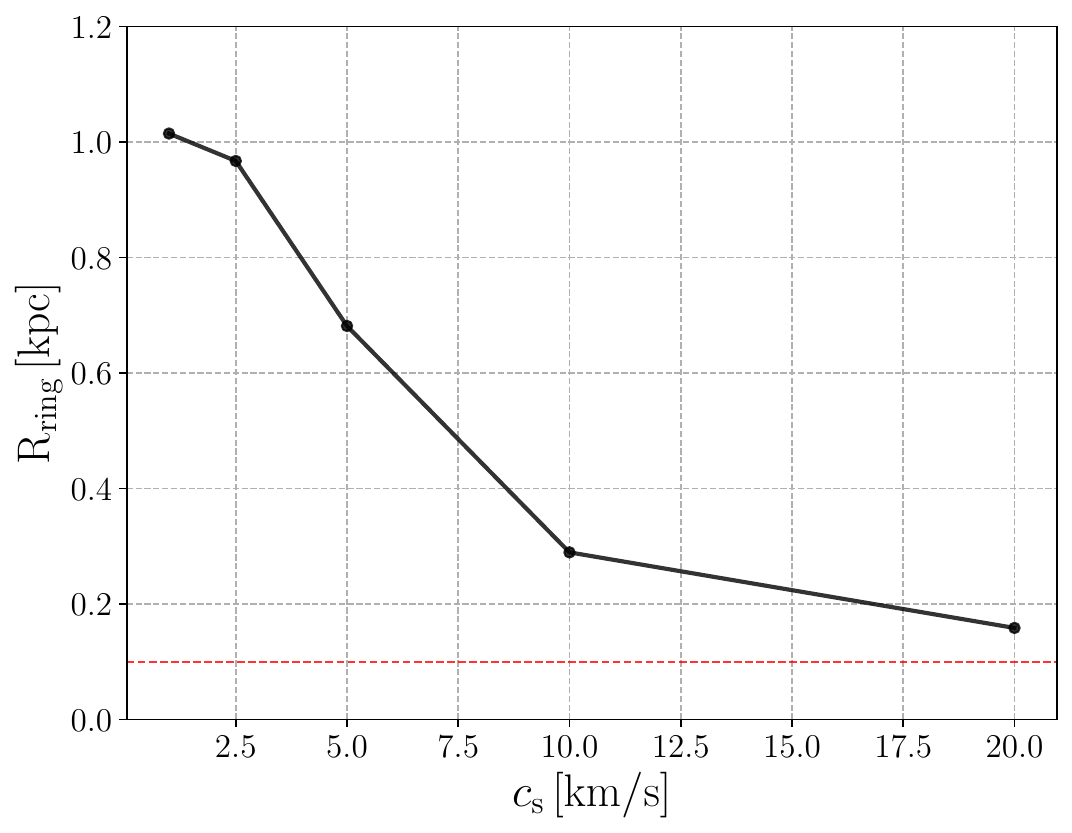}
    \caption{\emph{Top:} The full lines show the radius of the ring as a function of time in the simulations 01-05. The radius of the ring is calculated as $R_{\rm ring}=\sqrt{R_x R_y}$, where $R_x$ and $R_y$ are the locations of the density maxima along the $x$ and $y$ axis respectively. The dashed lines show the prediction according to Eq.~\eqref{eq:evo} obtained in the linear approximation (see Sect.~\ref{sec:secondstage}). \emph{Bottom:} The radius of the ring mediated over simulation time $t=1152\mhyphen1252\Myr$ as a function of the sound speed. The radius strongly depends on the sound speed. The red dashed line indicates the inner boundary of the computational grid.}
    \label{fig:ringsize}
\end{figure}

The final size of the ring depends very strongly on the sound speed (rightmost column in Fig.~\ref{fig:manyrho}). This is further quantified in Fig.~\ref{fig:ringsize}, which shows the evolution of the ring size as a function of sound speed. As can be seen in the bottom panel, increasing the sound speed by a factor of two can change the final ring size by the same factor.

\subsection{Disc with initial radius larger than the ILR} \label{sec:numericalresults2}

Simulations 04 and 04\_Large only differ in the size of the initial gas disc, $R_{\rm disc}=1.2\kpc$ vs.\ $R_{\rm disc}=5\kpc$. Thus, simulation 04 includes only the flow inside the ILR ($R_{\rm ILR}=1.61\kpc)$, while simulation 04\_Large comprises the large-scale flow in the entire bar region.

\begin{figure*}
	\includegraphics[width=\textwidth]{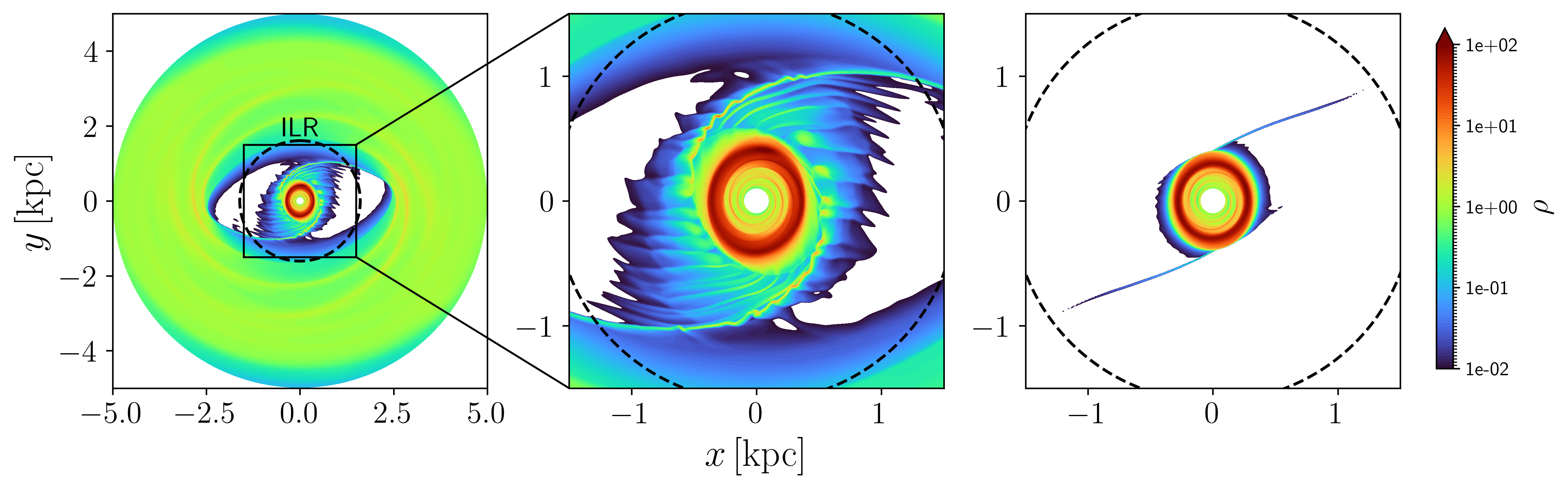}
    \caption{Surface density of the 04\_Large (left and zoom-in panel) and 04 (right) simulation at the end of the simulation ($t=1252\Myr$). The only difference between the two simulations is that in simulation 04\_Large the initial gas disc extends to $R_{\rm max}=5.0\kpc$, while in simulation 04 only to $R_{\rm disc}=1.2\kpc$. Simulation 04 is the same as shown in the second row of Fig.~\ref{fig:manyrho}. The dashed circle indicates the ILR. All panels are rotated so that the major axis of the bar potential (i.e. the $\theta=0$ line in Eq.~\ref{eq:potential}) coincides with the $x$ axis. The sense of rotation is clockwise. Comparison between the two simulations shows that the ring always reaches the same final size, regardless of the larger-scale flow outside the ILR, demonstrating that the physical process determining the radius of the ring must be ``local''.
    }
    \label{fig:fullvssmall}
\end{figure*}

Figure~\ref{fig:fullvssmall} shows that the final ring size is approximately the same in both simulations, and therefore that ring size does not depend on the large-scale flow outside the ILR. This implies that the mechanism determining the radius of the ring must be ``local'' (see point 5 in Sect.~\ref{sec:constraints}).

\begin{figure}
	\includegraphics[width=\columnwidth]{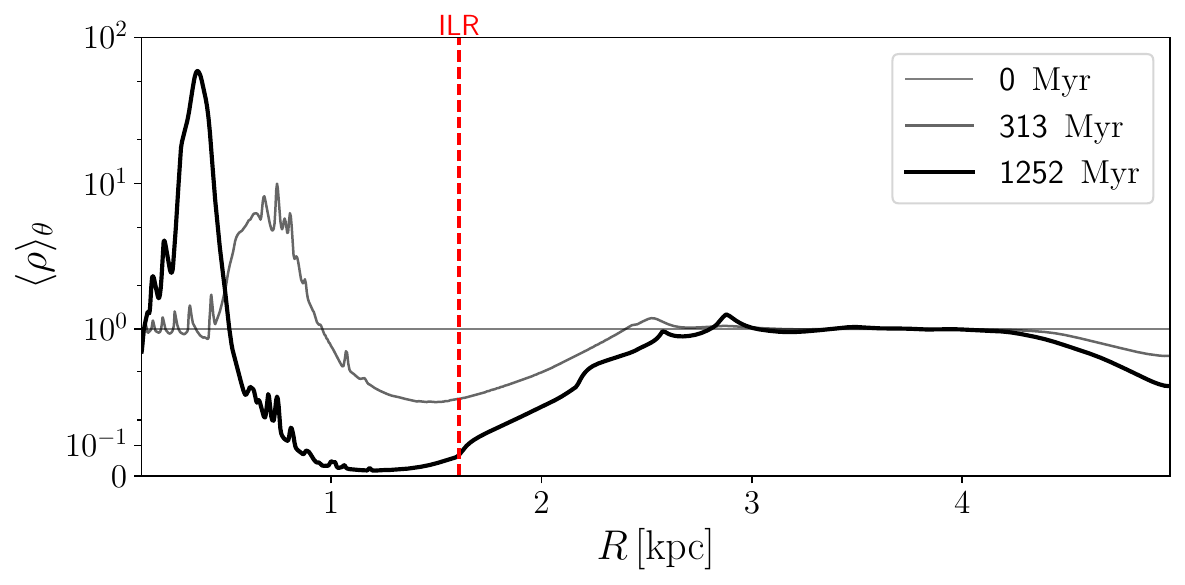}
    \caption{Axisymmetrised surface density $\langle \rho \rangle_\theta = \int \rho(R,\theta) \di \theta / (2 \pi)$ as a function of cylindrical radius $R$ for the simulation 04\_Large at three different times. An extensive gap opens around the ILR. The material that once was in the gap is transported inwards and accumulates at the inner edge of the gap, forming a nuclear ring.}
    \label{fig:sigma}
\end{figure}

Figure~\ref{fig:sigma} illustrates the evolution of the axisymmetrised surface density as a function of radius in the simulation 04\_Large. In particular, we can see that an extensive gap of low surface density is opened around the ILR. The nuclear ring is the inner edge of this gap, where the material that once was in the gap has accumulated.

\begin{table}
    \caption{Summary of the simulations run in this paper. The parameters are defined in Sect.~\ref{sec:numericalexperiments}.}
	\centering
	\begin{tabular}{lllll} 
		ID  & $\cs$ & $R_{\rm max}$  & $R_{\rm disc} $ &  $N_R$ \\
		 &  [$\kms$] &  [\kpc] & [\kpc] \\
		\hline
         01 & 1  & 1.5  & 1.2  & 512 \\
         02 & 2.5 & 1.5  & 1.2  & 512 \\
         03 & 5  & 1.5  & 1.2  & 512 \\
         04 & 10  & 1.5  & 1.2 & 512 \\
         05 & 20  & 1.5  & 1.2  & 512 \\
         04\_Large & 10 & 5.0  & 5.0 & 740 \\
         \hline
	\end{tabular}
\label{tab:sims}
\end{table}

\section{Previous theories, and what we look for in a theory} \label{sec:previous}

\subsection{Conditions that a plausible theory must satisfy} \label{sec:constraints}

We introduce the conditions that we believe any plausible theory for the formation of nuclear rings must satisfy. We take the approach that numerical experiments, such as those in Sect.~\ref{sec:numericalexperiments}, guide us on how the ring properties should depend on the underlying parameters. We summarise the insights obtained from simulations into the following five conditions:

\begin{itemize}
    \item[\bf 1.] \emph{The radius of the ring must depend on the circular rotation curve}. \cite{Athan92b} and \cite{Li2015} have shown that the radius of the ring in simulations changes if we change the circular velocity curve of the underlying gravitational potential, i.e.\ if we change the axisymmetric part of the gravitational potential (see in particular Fig.~4 in \citealt{Li2015}), while keeping everything else fixed.
    \item[\bf 2.] \emph{The radius of the ring must depend on the non-axisymmetric part of the underlying potential}. \cite{Sormani2015c} has shown that the radius of the ring can change significantly if we change the quadrupole of the potential while keeping the monopole (and therefore the rotation curve) fixed. Hence, a theory aiming to explain why rings form at a certain location must take into account a dependence on the non-axisymmetric part of the potential.
    \item[\bf 3.] \emph{The radius of the ring must depend on the bar pattern speed}. Many authors (e.g.\ \citealt{Athan92b,Li2015,Sormani2015c}, among others) have shown that the radius of the ring depends on the rotation speed of the bar. 
    \item[\bf 4.] \emph{The radius of the ring must depend on the equation of state of the gas.} Many authors (e.g.\ \citealt{Englmaier1997,Patsis2000,Kim++2012a,Sormani2015a}, among many others) have shown that the size of the ring strongly depends on the sound speed. This is confirmed by the numerical experiments we conducted in Section~\ref{sec:numericalresults1} (see in particular Figs.~\ref{fig:manyrho} and \ref{fig:ringsize}). Thus, the radius of the ring does not depend solely on the gravitational potential but must involve the equation of state of the gas. 
    \item[\bf 5.] \emph{The radius of the ring must be determined ``locally''}, i.e. the final ring size should not depend on the larger-scale flow at $R>R_{\rm ILR}$. This is demonstrated by the numerical experiments in Fig.~\ref{fig:fullvssmall}, which shows two simulations that differ only for the extent of the simulated gas disc. The 04\_Large simulation (left) covers the entire ``bar region'', out to $R_{\rm disc}=5\kpc$. It includes the usual bar-driven accretion flow from the disc to the ring. The 04 simulation (right) is the same shown in the second row of Fig.~\ref{fig:manyrho}, and it only simulates a gas disc of $R_{\rm disc}=1.2\kpc$, which is all contained within the ILR at $R_{\rm ILR}=1.61 \kpc$ (Fig.~\ref{fig:vcirc}). The final size of the ring is essentially the same in the two simulations (it is slightly larger in the 04\_Large simulation because fresh gas is continuously brought from outside the ILR, which takes longer to lose angular momentum). This shows that removal of angular momentum also happens in the vicinity of the ring. Material from outside the ILR that crosses the ILR continually loses angular momentum up to when it settles on the ring.
\end{itemize}

\subsection{Previous theories for the formation of nuclear rings} \label{sec:previoustheories}

Here we briefly summarise previous theories for the formation of the ring and for determining its location. We argue that none of them provides a satisfactory explanation for the formation of the rings by showing that each of them fails to satisfy at least one of the conditions outlined in Section~\ref{sec:constraints}.

\begin{itemize}
    \item \emph{The resonant theory \citep{Combes1988,Combes1996,Buta1996}}. This is perhaps the most widely accepted theory, especially in the extragalactic community. It states that the ring forms at the Lindblad resonance under the continuous action of gravity torques from the bar potential.
    
    \item[$\boldsymbol{\times}$] \emph{Refutation}: This theory satisfies conditions 1, 3, 5, and if the notion of ILR is generalised to include strongly barred potentials \citep{vanAlbada1982,Athan92a}, it may satisfy condition 2. However, since the position of the resonance does not depend on the equation of state of the gas, it does not satisfy condition 4. Moreover, the numerical experiments shown in Fig.~\ref{fig:manyrho} show that the radius of the ring forms at a radius $R$ that is much smaller than $R_{\rm ILR}=1.61\kpc$ (on this point, see also \citealt{Regan2003}).
    
    \item \emph{The minimum shear theory \citep{Lesch1990,KrumholzKruijssen2015}}. This theory states that the ring forms at the radius at which the shear, as calculated from the axisymmetric rotation curve, is minimum. This conclusion stems from an analogy with accretion disc theory, in which transport is more efficient where shear is higher, so gas is expected to pile up and form a ring at the point of minimum shear.

    \item[$\boldsymbol{\times}$]  \emph{Refutation}: This theory satisfies conditions 1 and 5, but does not satisfy conditions 2, 3, 4. \cite{SormaniLi2020} demonstrate in detail that simulations are inconsistent with this theory.

    \item \emph{The reverse shear theory \citep{Sormani+2018b}}. This theory states that the ring forms in a region where a family of non-axisymmetric closed periodic orbits called $x_2$ orbits displays ``reverse shear'' that prevents viscous spreading, making it possible to confine a stable ring.

   \item[$\boldsymbol{\times}$] \emph{Refutation}: This theory satisfies conditions 1, 2, 3, and 5, but it does not satisfy condition 4 since it predicts that the radius of the ring depends exclusively on the gravitational potential.
    
\end{itemize}
In conclusion, none of the currently available theories satisfies all the criteria introduced in Sect.~\ref{sec:constraints}, highlighting the need for a new theory.

\section{Linear disc dynamics} \label{sec:lineardisc}

The simulations in Sect.~\ref{sec:numericalexperiments} suggest that density waves are important for the removal of angular momentum of the disc and the opening of the gap. To gain insight into this process, in this section we study the excitation of waves by the bar potential in the linear regime, and estimate the amount of angular momentum that they remove from the gas disc as a function of the unperturbed density profile $\rho_0$ and of the sound speed $\cs$. As we shall see, the excitation of waves happens primarily at two locations: i) near the ILR. This regime has been studied in detail by \cite{Goldreich1979}, and we will not repeat their calculations here; ii) at sharp edges (i.e., strong gradients) in the unperturbed density distribution $\rho_0$. The insight gained in this section will be used in Sect.~\ref{sec:formation} to develop a new picture for the formation of nuclear rings.

Consider a 2D axisymmetric differentially rotating fluid disc in equilibrium in an external gravitational potential. Our goal is to study the propagation of small perturbations (waves) and their excitation by a ``small'' external potential. These ``density waves'' are conceptually similar to sound waves in air, but the rotation makes the dynamics richer and mathematically more complex. In Appendix~\ref{appendix:toymodel} we present a 1D toy problem that can be solved fully analytically and provides a mathematically simpler analog to the more complicated problem studied in this section.

We start from the disc's unperturbed steady state and linearize the equations of motions around it. We ignore the self-gravity of the gas. The equations of motion are the continuity and Euler equations:
\begin{align} 
& \pa_t \rho + \nabla \cdot (\rho \bfv) = 0 \,, 								 \label{eq:c01} \\
 & \pa_t \bfv + (\bfv \cdot \nabla) \bfv = - \frac{ \nabla P}{\rho}  - \nabla \Phi \,,	\label{eq:e01}
\end{align}
where $\rho$ is the surface density, $\bfv=v_x \hatex + v_y \hatey$ is the velocity, $P$ is the pressure, and $\Phi(\bfx,t)$ is the external gravitational potential. We assume a polytropic equation of state
\begin{equation} \label{eq:eos}
P = K \rho^\gamma,
\end{equation}
where $\gamma\geq1$ and $K$ is a constant. To simplify the calculations it is convenient to introduce the enthalpy $h$ defined by:
\begin{equation} \label{eq:enthalpy}
\nabla h = \frac{ \nabla P}{\rho} \,,
\end{equation}
substituting \eqref{eq:eos} into \eqref{eq:enthalpy} and integrating we find:
\begin{equation}  \label{eq:enthalpy2}
h = \begin{cases} K \left( \frac{\gamma}{\gamma-1} \right) \rho^{\gamma-1} \qquad & \text{if} \quad \gamma>1	\,, \\  K  \log \rho \qquad & \text{if} \quad \gamma = 1 \,.\end{cases}
\end{equation}
Using \eqref{eq:enthalpy}, the equations of motion \eqref{eq:c01} and \eqref{eq:e01} can be expanded in polar coordinates $(R,\theta)$ as:
\begin{align} 
& \pa_t \rho + \frac{1}{R} \pa_R \left(R \rho v_R \right) + \frac{1}{R} \pa_\theta \left( \rho v_\theta \right) = 0 \,, 								 \label{eq:c02} \\
& \pa_t v_R + \left( v_R \pa_R + \frac{v_\theta}{R} \pa_\theta \right) v_R - \frac{v_\theta^2}{R} = - \pa_R h  - \pa_R \Phi \,,			 \label{eq:e02R} \\
 & \pa_t v_\theta + \left( v_R \pa_R + \frac{v_\theta}{R} \pa_\theta \right) v_\theta + \frac{v_R v_\theta}{R} = -  \frac{1}{R}  \pa_\theta h  -  \frac{1}{R}  \pa_\theta \Phi \,.			 \label{eq:e02theta}
\end{align}

\subsection{Unperturbed state} \label{sec:unperturbed}

We assume that the density, velocity and gravitational potential of the unperturbed steady-state are:
\begin{align}
\rho & = \rho_0(R), \\
h & = h_0(R), \\
\bfv & = \Omega(R) R \, \hatetheta, \\
\Phi & = \Phi_0(R).
\end{align}
Substituting these into \eqref{eq:c02}-\eqref{eq:e02theta} and assuming steady-state and axisymmetry ($\pa_t = \pa_\theta = 0$), we see that the continuity equation \eqref{eq:c02} and the azimuthal Euler equation \eqref{eq:e02theta} are already satisfied, while the radial Euler equation \eqref{eq:e02R} gives:
\begin{align} 
\Omega^2 R = \frac{ \di (h_0 + \Phi_0)}{\di R}. \label{eq:unperturbed}
\end{align}
In the following, $h_0$, $\Phi_0$ and $\Omega$ are prescribed functions of $R$ that satisfy Eq.~\eqref{eq:unperturbed}. Note that given $\Phi_0(R)$, there formally exists an equilibrium solution $h_0(R)$ for any arbitrary rotation profile $\Omega(R)$. However not all possible profiles are physical. To avoid instability, the unperturbed state must satisfy the Rayleigh stability criterion, which states that a necessary and sufficient condition for the local axisymmetric stability of an inviscid differentially rotating fluid disc is that the specific angular momentum monotonically increases with $R$, i.e.\footnote{The Rayleigh criterion is equivalent to the condition that the epicyclic frequency is real ($\kappa^2>0$, see Eqs.~\ref{eq:OortB} and \ref{eq:epyciclic}).}
\begin{equation} \label{eq:Rayleigh}
    \frac{\di(R^2 \Omega)}{\di R}>0 \,. \qquad \text{(Rayleigh criterion)}
\end{equation}
In this paper we will assume mainly two types of density profiles. The first is a constant density profile 
\begin{equation}
    \rho_0(R)=\bar{\rho}=\rm constant\,.
\end{equation}
The second is a truncated disc profile, i.e. a density that is roughly constant at $R\ll R_{\rm edge}$, has a relatively sharp transition at an edge $R_{\rm edge}$ during which it drops at a much lower value, and is then roughly constant again at $R>R_{\rm edge}$. Note that the edge cannot be made too thin, otherwise it would violate the Rayleigh criterion \eqref{eq:Rayleigh}. When later in the paper it will be necessary to assume a specific truncated profile for numerical calculations, we will use the following:
\begin{equation} \label{eq:edge}
   \rho_0(R) = \frac{\bar{\rho}}{2} \left[1 - \frac{a}{\left(1+a^2\right)^{1/2}} \right]\,,
\end{equation}
where
\begin{equation}
 a = \frac{R - R_{\rm edge}}{\Delta R} \,,
\end{equation}
where $R_{\rm edge}$ is the position of the edge and $\Delta R$ controls its width. The quantity $\bar{\rho}$ is a constant that, as noted in Sect.~\ref{sec:numerics}, essentially defines the units used for density. Physically meaningful results do not depend on the particular value of this quantity since the equations of motion \eqref{eq:c01} and \eqref{eq:e01} are invariant under density rescaling. Without loss of generality, we set $\bar{\rho}=1$.

\subsection{Linearised equations}

To study the propagation of small waves on top of the unperturbed state described in the previous section, we expand all quantities as:
\begin{align}
\rho & = \rho_0 + \rho_1 	\,,	\label{eq:rho1} 	\\
h     & = h_0 + h_1 		\,, 	\label{eq:h1}	\\
\bfv & = \bfv_0 + \bfv_1 	\,, 	\label{eq:v1}	\\
\Phi & = \Phi_0 + \Phi_1 	\,.	\label{eq:Phi1}
\end{align}
Substituting equations~\eqref{eq:rho1}-\eqref{eq:Phi1} into \eqref{eq:c02}-\eqref{eq:e02theta} and linearising by keeping only first-order terms in the quantities with subscript 1, we obtain:
\begin{align}
& \frac{D}{Dt}\left(\frac{\rho_1}{\rho_0} \right) + \frac{\di \log(R \rho_0)}{\di R}  v_{R1} + ( \pa_R v_{R1} ) + \frac{1}{R} (\pa_\theta v_{\theta 1}) = 0 \,, \label{eq:linear01a} \\
& \frac{D v_{R1}}{Dt} - 2 \Omega v_{\theta 1} = - \pa_R \left [ h_1 + \Phi_1 \right] \,, \label{eq:linear01b}\\
& \frac{D v_{\theta1}}{Dt} + 2 B v_{R1} = - \frac{1}{R} \pa_\theta \left [ h_1 + \Phi_1 \right] \,, \label{eq:linear01c}
\end{align}
where we have defined the convective derivative of the unperturbed state
\begin{equation}
    \frac{D}{Dt} = \pa_t + \Omega \pa_\theta \,,
\end{equation}
 and the Oort parameter
\begin{equation} \label{eq:OortB}
    B(R) = \Omega + \frac{R}{2} \frac{\di \Omega}{\di R} \,.
\end{equation}
Without loss of generality, we can write all the ``small'' subscript-1 quantities as:
\begin{align}
\rho_1 & = \tilde{\rho}_1(R) \exp [ i (m \theta - \omega t)], \label{eq:lin02a} \\
v_{1R} & = \tilde{v}_{1R}(R) \exp [ i (m \theta - \omega t)], \\
v_{1\theta} & = \tilde{v}_{1\theta}(R) \exp [ i (m \theta - \omega t)], \\
h_{1} & = \tilde{h}_{1}(R) \exp [ i (m \theta - \omega t)], \\
\Phi_{1} & = \tilde{\Phi}_{1}(R) \exp [ i (m \theta - \omega t)], \label{eq:lin02e}
\end{align}
where $\tilde{\rho}_1$, $\tilde{v}_{1R}$ etc. are complex, and the ``physical'' quantity is the real part. The general solution of equations~\eqref{eq:linear01a}-\eqref{eq:linear01c} can always be decomposed in such modes because the equations are linear and the superposition principle applies. Each mode evolves independently from the others in the linear approximation. In this paper, we will only be concerned with $m=2$ as this is the only non-zero term in the expansion of the external potential described in Appendix~\ref{appendix:potential}. Hereafter, we drop the $\,\tilde{}\,$ symbol to avoid cluttering. With these substitutions, we have $\pa_t = -i \omega$ and $\pa_\theta=i m$. Substituting Eq.~\eqref{eq:lin02a} into \eqref{eq:enthalpy2} we have:
\begin{equation} \label{eq:enthalpy3}
    h_1 = \cs^2 \left( \frac{\rho_1}{\rho_0} \right)
\end{equation}
where we have introduced the sound speed of the unperturbed medium:
\begin{equation} \label{eq:cs}
    \cs^2 = \gamma K \rho_0^{\gamma-1}\,.
\end{equation}
Eq.~\eqref{eq:enthalpy3} is valid for $\gamma \geq 1$ (including equality). We also define
\begin{equation}
    \Omegap = \frac{\omega}{m} \,.
\end{equation}
This is the angular frequency with which each mode appears to rotate, as can be understood by noting that
\begin{equation}
    e^{i(m\theta-\omega t)}=e^{im(\theta - \Omegap t)} \,.
\end{equation}
 In this paper, we will always take $\Omegap$ to be the same as the pattern speed of the bar described in Appendix~\ref{appendix:potential}, since only modes at this frequency can be excited by the external potential in the linear approximation.

Substituting \eqref{eq:lin02a}-\eqref{eq:lin02e} into \eqref{eq:linear01a}-\eqref{eq:linear01c} we obtain:
\begin{align}
& i m \left(\Omega - \Omegap\right) \left(\frac{\rho_1}{\rho_0} \right) + \frac{\di \log(R \rho_0)}{\di R}  v_{R1} +  \frac{\di v_{R1}}{\di R}  + \frac{i m}{R}  v_{\theta 1} = 0 \label{eq:c03} \,, \\
& i m \left(\Omega - \Omegap\right) v_{R1} - 2 \Omega v_{\theta 1} = - \frac{\di}{\di R} \left [ h_1 + \Phi_1 \right] \,, \label{eq:lin03a} \\
& i m \left(\Omega - \Omegap\right) v_{\theta1} + 2 B v_{R1} = - \frac{i m}{R}  \left [ h_1 + \Phi_1 \right] \label{eq:lin03b} \,,
\end{align}
Isolating $v_{R1}$ and $v_{\theta 1}$ from \eqref{eq:lin03a} and \eqref{eq:lin03b} we find:
\begin{align}
& v_{R1} = - \frac{im}{D} \left(   \frac{2 \Omega}{R} + \left(\Omega - \Omega_{\rm p} \right)  \frac{ \di }{\di R}  \right)  \left[ h_1 + \Phi_1 \right] \label{eq:vR1} \\
& v_{\theta1} = \frac{1}{D} \left( \frac{m^2(\Omega-\Omega_{\rm p})}{R} + 2 B \frac{\di}{\di R} \right)  \left[ h_1 + \Phi_1 \right] \label{eq:vtheta1}
\end{align}
where we have defined
\begin{align}
    D &  = \kappa^2 - m^2 (\Omega - \Omegap)^2, \label{eq:D}  \\
    \kappa^2 & = 4 B \Omega, \qquad \text{(epicyclic frequency)}. \label{eq:epyciclic}
\end{align}
The points where $D=0$ define the Lindblad resonances,\footnote{Note that when $
\rho_0$ is not constant, the position where $D=0$ can differ slightly from the value of the ILR given in Appendix~\ref{appendix:potential} because of the contribution from the pressure term $h_0$ to $\Omega$ in Eq.~\eqref{eq:unperturbed}.} while the point where $\Omega=\Omegap$ defines the Corotation resonance. Now we can substitute \eqref{eq:vR1} and \eqref{eq:vtheta1} into \eqref{eq:c03} and use \eqref{eq:enthalpy3} to eliminate $\rho_1$ to obtain an equation in the variable $h_1$:
\begin{equation} \label{eq:mainode}
\frac{\di^2 h_1}{\di R^2} + 2 H(R) \frac{\di h_1}{\di R} + W(R) h_1 = F(R)
\end{equation}
where
\begin{align}
H(R) & = \frac{1}{2}\frac{\di}{\di R} \left[ \log \left(  \frac{R \rho_0 }{D} \right) \right]\,,  \label{eq:H} \\
W(R) & =  C(R) -  \frac{D(R)}{\cs^2}\,, \label{eq:W} \\
C(R) & =  \left( \frac{ 2 \Omega}{R ( \Omega - \Omega_{\rm p})} \right)   \frac{\di}{\di R}  \left[ \log{ \left( \frac{\rho_0\Omega}{D} \right) } \right] - \frac{ m^2}{R^2}\,,  \label{eq:C} \\
F(R) & = - \left\{   \frac{\di^2}{\di R^2}  +  2 H(R) \frac{\di}{\di R}  + C(R) \right\} \Phi_1(R) \,. \label{eq:F2}
\end{align}
Eq.~\eqref{eq:mainode}
coincides with Eq.~(13) of \citet{Goldreich1979}. The same equation has been also derived by others \citep[e.g.][]{Feldman1973,Bertin1989}. It is a second order ordinary differential equation with non-constant coefficients $H(R)$ and $W(R)$. The term $F(R)$ is a forcing term (recall that $\Phi_1(R)$ is externally prescribed). Note that $H(R)$ and $W(R)$ diverge where $(\Omega-\Omegap)=0$ and where $D=0$, i.e. at the corotation and Lindblad resonances.

In order to eliminate the first order derivative from Eq.~\eqref{eq:mainode}, it is convenient to define a new variable $g_1$ such that
\begin{equation}
\label{eq:g1}
h_1 = \left(\frac{|D|}{R\rho_0}\right)^{1/2}g_1\;.
\end{equation}
Substituting Eq.~\eqref{eq:g1} into Eq.~\eqref{eq:mainode}, one finds
\begin{equation}
\label{eq:mainode2}
\boxed{\frac{\di^2 g_1}{\di R^2} + K^2(R) g_1 = Q(R)} \;,
\end{equation}
where
\begin{align} \label{eq:dispersionrelation}
K(R) & = \left[W-H^2-\frac{\di H}{\di R}\right]^{1/2}, \\
Q(R) & = \left(\frac{R\rho_0}{|D|}\right)^{1/2} F(R) \;.
\end{align}
Eq.~\eqref{eq:mainode2} is the fundamental equation that governs linear modes in the disc. It is similar to Eq.~\eqref{eq:exc_100} in the toy problem in Appendix~\ref{appendix:toymodel}, but is more complicated because $K$ is not constant. To follow the calculations in the following section more easily, it is useful to note that Eq.~\eqref{eq:mainode2} is equivalent to that of a forced harmonic oscillator, $m \ddot{x} + k^2(t) x = q(t)$, where $t$ replaces $R$, $m$ is the mass, $k(t)$ is a time-dependent spring constant, and $q(t)$ is a time-dependent external force.

\begin{figure}
	\includegraphics[width=\columnwidth]{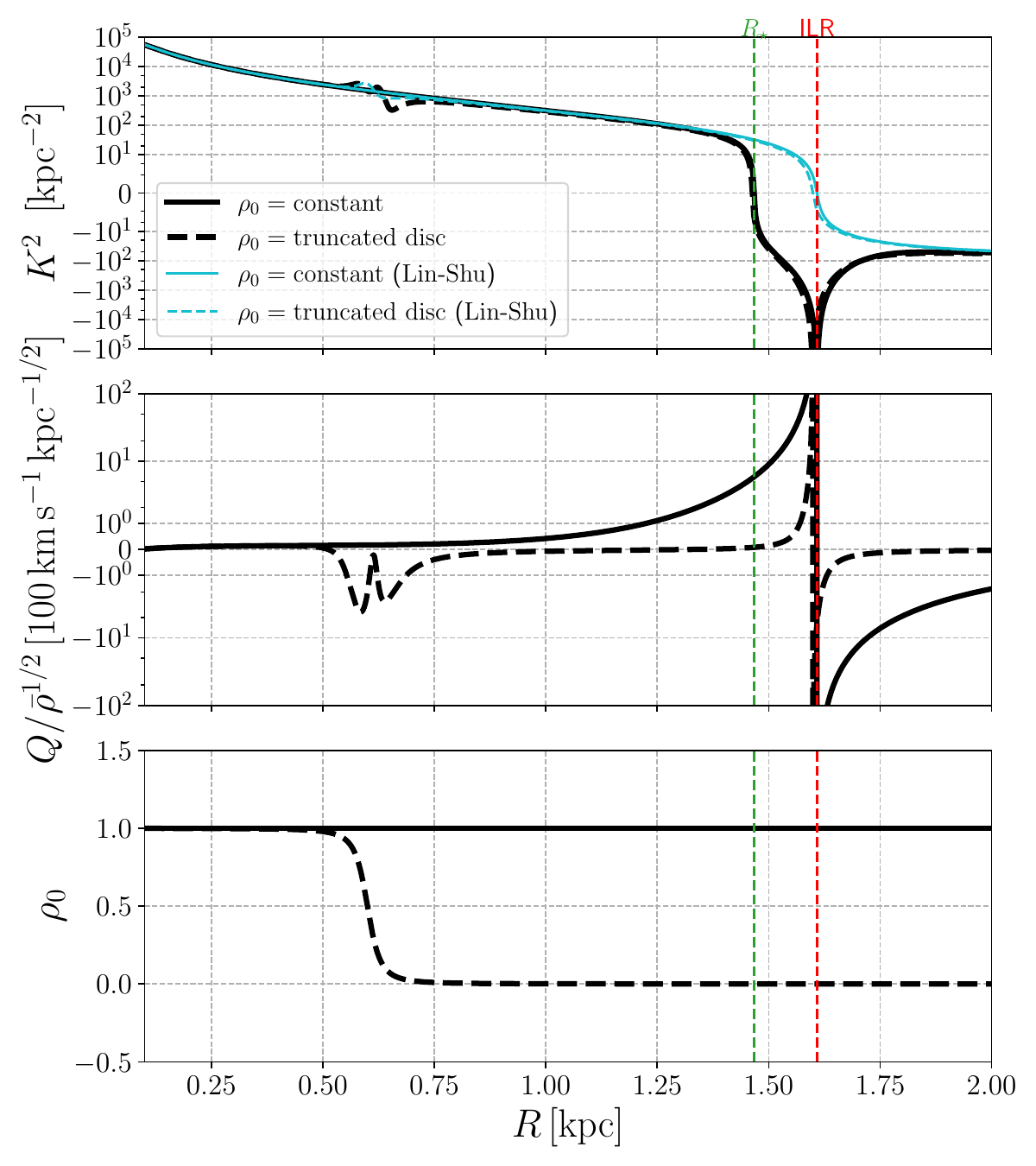}
    \caption{\emph{Top}: the coefficient $K(R)$ in Eq.~\eqref{eq:mainode2} for a uniform disc (full black line) and a truncated disc (dashed black line) in the case $\cs=10\kms$. In the WKB approximation, this represents the wavenumber versus radius of free density waves that rotate with the same pattern speed of the bar (see Sect.~\ref{sec:WKB}). The cyan lines compare with the wavenumber given by the Lin-Shu dispersion relation \eqref{eq:KLS}. \emph{Middle}: the forcing term $Q(R)$ in Eq.~\eqref{eq:mainode2}. \emph{Bottom}: the uniform ($\rho_0=1$) and truncated disc (Eq.~\ref{eq:edge} with $R_{\rm edge}=0.6\kpc$ and $\Delta R = 0.03\kpc$) density profiles assumed in this figure. The red vertical dashed line marks the ILR. The green vertical dashed line marks $R_\star$, which is defined as the radius where $K(R)=0$ (see Sect.~\ref{sec:Rstar}).}
    \label{fig:coefficients}
\end{figure}

\subsection{Analysis of Eq.~\eqref{eq:mainode2}} \label{sec:mainodeanalysis}

\begin{figure}
	\includegraphics[width=\columnwidth]{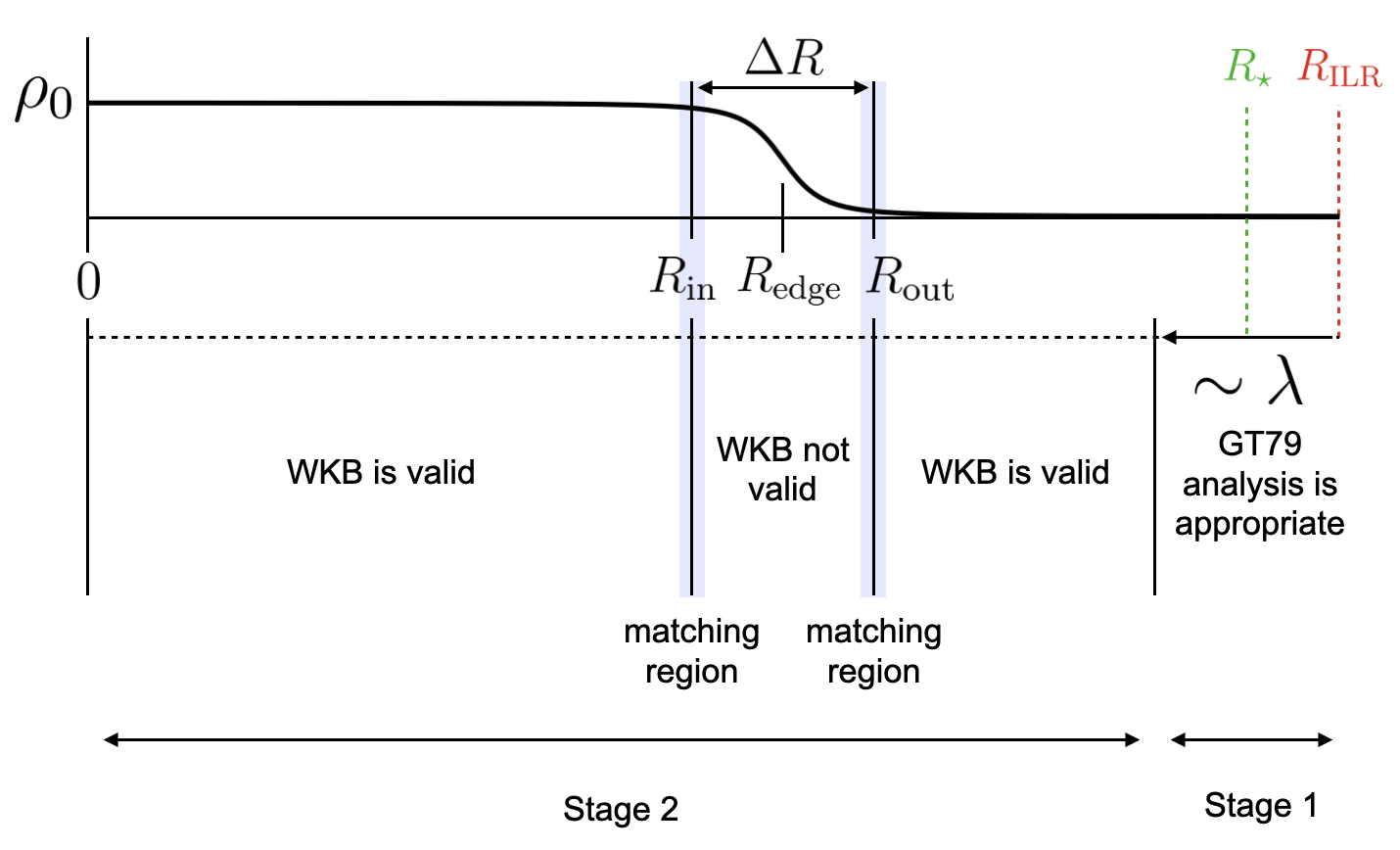}
    \caption{Schematic diagram of where the various approximate solutions of Eq.~\eqref{eq:mainode2} apply. ``WKB is valid'' denotes where the general solution of Eq.~\eqref{eq:mainode2} is well approximated as the sum of the WKB solution \eqref{eq:WKB1} and the equilibrium solution $g_Q$ given by \eqref{eq:gQ}. ``WKB not valid'' denotes the region near the edge where Eq.~\eqref{eq:g1neargen} is more appropriate. The shaded ``matching regions'' denote where both solutions are simultaneously valid and we can apply the method of matched asymptotic expansions. The region within approximately one wavelength  $\lambda$ from the ILR is where the analysis of \citet{Goldreich1979} is appropriate. ``Stage 1'' and ``Stage 2'' denote the regions correponding to the two stages in our picture of the formation of the rings described in Sect.~\ref{sec:formation}.}
    \label{fig:schematic}
\end{figure}

Equation~\eqref{eq:mainode2} describes the dynamics of the most general linear perturbation in the presence of an external potential. To calculate the amplitude of waves excited by the bar potential we need to solve this equation with appropriate boundary conditions. Unfortunately, no general analytic solution is available, so we need to resort to various approximations that are valid in different radial ranges. Figure~\ref{fig:schematic} provides an overview of the various regimes that we analyse.

This section is structured as follows. In Sect.~\ref{sec:Rstar} we identify special points where the treatment of Eq.~\eqref{eq:mainode2} require special care because the coefficient $K$ either vanishes or diverges. In Sect.~\ref{sec:WKB} we derive the WKB solution of the homogeneous equation associated with \eqref{eq:mainode2}, and show that it is generally very accurate away from the special points and away from sharp edges (see Fig.~\ref{fig:schematic}). In Sect.~\ref{sec:equilibriumsol} we derive a particular solution of the non-homogeneous \eqref{eq:mainode2} that is approximately valid when $\cs$ is sufficiently low and away from special points and sharp edges. In Sect.~\ref{sec:excitation} we obtain exact numerical solutions of Eq.~\eqref{eq:mainode2} in a few selected cases, to illustrate that truncated discs with sharp edges excite much stronger waves than uniform discs. In Sect.~\ref{sec:analyticedge} we present an approximated analytical solution of Eq.~\eqref{eq:mainode2} that is valid near sharp edges and estimate the flux of angular momentum at sharp edges.

\subsubsection{Special points} \label{sec:Rstar}

There are two types of points where Eq.~\eqref{eq:mainode2} requires special attention:
\begin{itemize}
\item \emph{Turning points}. These are the points $R_\star$ where $K(R_\star)=0$. At these points, the character of the solutions changes from oscillatory to exponential.
\item \emph{Singular points}. These are points where $K(R)$ diverges. As can be seen from Eqs.~\eqref{eq:dispersionrelation} and \eqref{eq:H}-\eqref{eq:F2}, this happens at the Lindblad and Corotation resonances.
\end{itemize} 
Figure~\ref{fig:coefficients} shows the coefficients of Eq.~\eqref{eq:mainode2} for a uniform and a truncated disc profile in the case $\cs=10\kms$. In the region of interest for this paper there is typically one turning point $R_\star$ and one singular point at $R_{\rm ILR}$, with $R_\star < R_{\rm ILR}$. As we shall see below, $R_\star$ is where the medium becomes absorbing and leading waves incident from $R<R_\star$ are reflected into trailing waves that subsequently travel inwards. The position of $R_\star$ depends on both the sound speed $\cs$ and the shape of the unperturbed density profile $\rho_0(R)$. In the limit $\cs\to0$ we have $R_\star \to R_{\rm ILR}$. However, for a finite value of the sound speed, the two points are distinct. 

\subsubsection{WKB solution of the homogeneous equation} \label{sec:WKB}

Consider the homogeneous equation associated with Eq.~\eqref{eq:mainode2}, i.e.\ the equation obtained setting $Q=0$. This equation describes the propagation of ``free'' density waves on top of the unperturbed disc in the absence of the external bar potential. In this case, Eq.~\eqref{eq:mainode2} is of the same form of Eq.~\eqref{eq:WKBex2_01} and it can be solved in the WKB approximation. The general solution is given by Eq.~\eqref{eq:WKBgensol}, which adapted to the notation used here reads:
\begin{align}
\label{eq:WKB1}
 g_1(R)  & = \frac{C_1}{\sqrt{K(R)}} \exp \left[i\int_{R_0}^R K(s) \di s \right] +\frac{C_2}{\sqrt{K(R)}} \exp \left[-i\int_{R_0}^R K(s) \di s \right]\,,
 \end{align}
where $C_1$ and $C_2$ are arbitrary complex constants and $R_0$ is an arbitrary radius.

The two terms on the right-hand side of Eq.~\eqref{eq:WKB1} represent two waves travelling in opposite directions, analogously to the two sound waves that are possible in a uniform medium at a given frequency (see Appendix~\ref{appendix:toymodel}). The quantity $K$ is the wavenumber, which varies with radius. When $K^2>0$, the solution \eqref{eq:WKB1} has oscillatory character and waves can travel, while when $K^2<0$ it has exponential character and the medium is absorbing. Thus, as can be seen from Fig.~\ref{fig:coefficients}, travelling waves can exist only at $R<R_\star$. Equation~\eqref{eq:dispersionrelation} implicitly contains $\omega$, and therefore for fixed $R$ this expression can be seen as a dispersion relation $K=K(\omega)$.

The direction of propagation of the waves can be understood from the group velocity. In Appendix~\ref{appendix:vg} we calculate the group velocity of the WKB waves and we find that trailing waves ($C_1\neq0$ and $C_2=0$) propagate inwards, while leading waves  ($C_1=0$ and $C_2\neq0$) propagate outwards. 

The angular momentum flux associated with the WKB waves \eqref{eq:WKB1} is calculated in Appendix~\ref{appendix:angularmomentum} and is given by Eq.~\eqref{eq:FAWKB}: 
\begin{equation} \label{eq:FAWKB2}
F_{\rm A} = m\pi \left(\left|C_1\right|^2 - \left|C_2\right|^2 \right)\;.
\end{equation}
Since $C_1$ and $C_2$ are constant for a given WKB wave, this equation shows that the flux of angular momentum is constant as a function of $R$. It can be shown that the angular momentum flux corresponds to the adiabatic invariant associated with the general WKB solution \eqref{eq:WKBgensol}. Equation~\eqref{eq:FAWKB2} also shows that the trailing wave has $F_A>0$, while the leading wave has $F_A<0$. Thus, trailing (leading) wave packets remove (increase) the amount of angular momentum in the region where they travel.

\begin{figure}
	\includegraphics[width=\columnwidth]{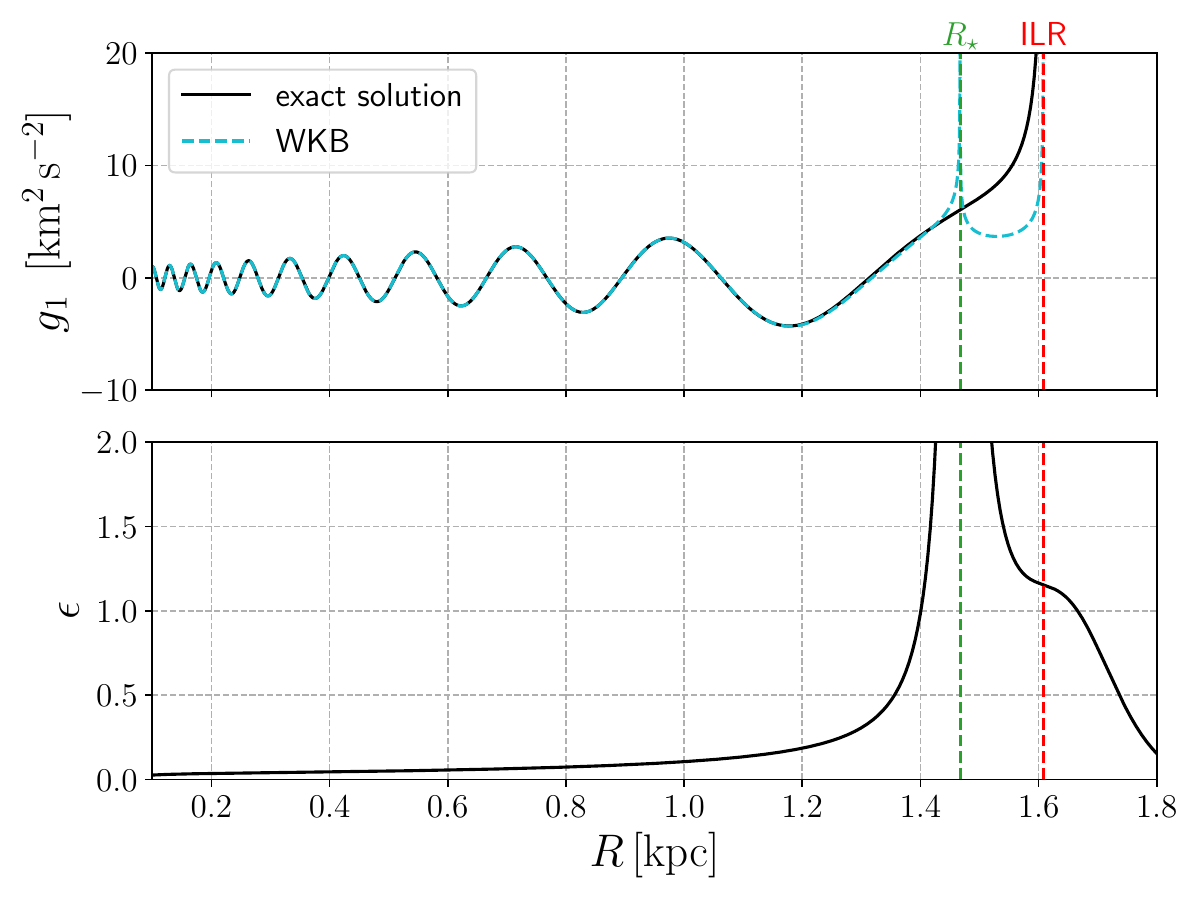}
    \caption{\emph{Top}: comparison between the exact solution of Eq.~\eqref{eq:mainode2} and the WKB approximation. The full black line shows the solution obtained numerically integrating Eq.~\eqref{eq:mainode2} from $R=0.1\kpc$ with initial conditions $g_1=1$, $\di g_1/ \di R=0$ (full black line). The cyan dashed line shows the WKB approximation \eqref{eq:WKB1}. We assumed $\cs=10\kms$ and constant unperturbed density $\rho_0(R)=1$. \emph{Bottom}: the ``small'' parameter of the WKB approximation (Eq.~\ref{eq:eps1}). At $R_\star$ it diverges. The WKB approximation works well only at $R<R_\star$.}
    \label{fig:WKBvsexact}
\end{figure}

What is the range of validity of the WKB approximation? The WKB approximation is expected to work well when the following parameter is small (see Eq.~\ref{eq:WKBcond}):
\begin{equation}
    \epsilon = \left| \frac{\di K /\di R}{K^2}  \right| \,, \label{eq:eps1} 
\end{equation}
Figure~\ref{fig:WKBvsexact} shows that the WKB approximation works exceptionally well at $R<R_\star$, but breaks down near $R=R_\star$. The WKB approximation will also fail near sharp edges, because $\di K/\di R$ becomes large (e.g.~Fig.~\ref{fig:coefficients}).

The WKB approximation used here is not completely equivalent to the more well-known Lin-Shu approximation. The Lin-Shu dispersion relation in the absence of self-gravity ($G=0$) is given by (Eq. 6.55 of \citealt{BT2008}):
\begin{equation} \label{eq:KLS}
K_{\rm Lin\mhyphen Shu}^2 = -\frac{D}{\cs^2} \,,
\end{equation}
where $D$ is given by Eq.~\eqref{eq:D}. The top panel in Fig.~\ref{fig:coefficients} compares the Lin-Shu dispersion relation (cyan line) with the dispersion relation given by Eq.~\eqref{eq:dispersionrelation}. The two are similar at $R<R_\star$, but differ considerably around $R_\star$ and $R_{\rm ILR}$. In particular, in the Lin-Shu approximation the turning point (which is the point where waves are absorbed) coincides with the ILR, while it is at a smaller radius ($R_\star$) according to Eq.~\eqref{eq:dispersionrelation}. This is because the Lin-Shu dispersion relation assumes very small sound speed, while the dispersion relation \eqref{eq:dispersionrelation} takes into account the effect of finite sound speed. Indeed, in the limit of vanishing sound speed we recover the Lin-Shu dispersion relation from our dispersion relation \eqref{eq:dispersionrelation}. This can be shown by noting that in this limit $W(R) \simeq - D/\cs^2$ (Eq.~\ref{eq:W}), while $H^2 \ll W$ and $\di H/\di R \ll W$ (Eq.~\ref{eq:dispersionrelation}).

\subsubsection{Approximate non-oscillatory solution of the non-homogeneous equation} \label{sec:equilibriumsol}

An approximate particular solution of Eq.~\eqref{eq:mainode2} is:
\begin{equation} \label{eq:gQ}
    g_{Q}(R) = \frac{Q}{K^2} \,.
\end{equation}
In the analogy with the harmonic oscillator, this solution corresponds to following the ``instantaneous'' equilibrium position of the oscillator as the external force slowly varies. It is expected to be valid when the ``force'' $Q(R)$ varies slowly enough compared to the frequency of the harmonic oscillator . More formally, one can substitute $g_1=g_{Q}$ in Eq.~\eqref{eq:mainode2}, and impose that the first term on the left-hand side is small, i.e.~$\di^2 g_1/\di R^2\ll K^2 g_1$. This gives the following condition:  
\begin{equation} \label{eq:condgQ}
\frac{\di^2}{\di R^2}\left(\frac{Q}{K^2}\right) \ll Q \,.
\end{equation} This condition is verified in particular at low sound speed, since $K \to \infty$ as $\cs \to 0$ at fixed $R$ (see Eq.~\ref{eq:dispersionrelation}).

Eq.~\eqref{eq:gQ} is equivalent to Eq.~(15) of \cite{Goldreich1979}. It is a non-wave solution which is the analogue of the black dashed solution for the toy problem in Fig.~\ref{fig:toy} in Appendix~\ref{appendix:toymodel}.

\subsubsection{Excitation of density waves in uniform and truncated discs} \label{sec:excitation}

\begin{figure}
	\includegraphics[width=\columnwidth]{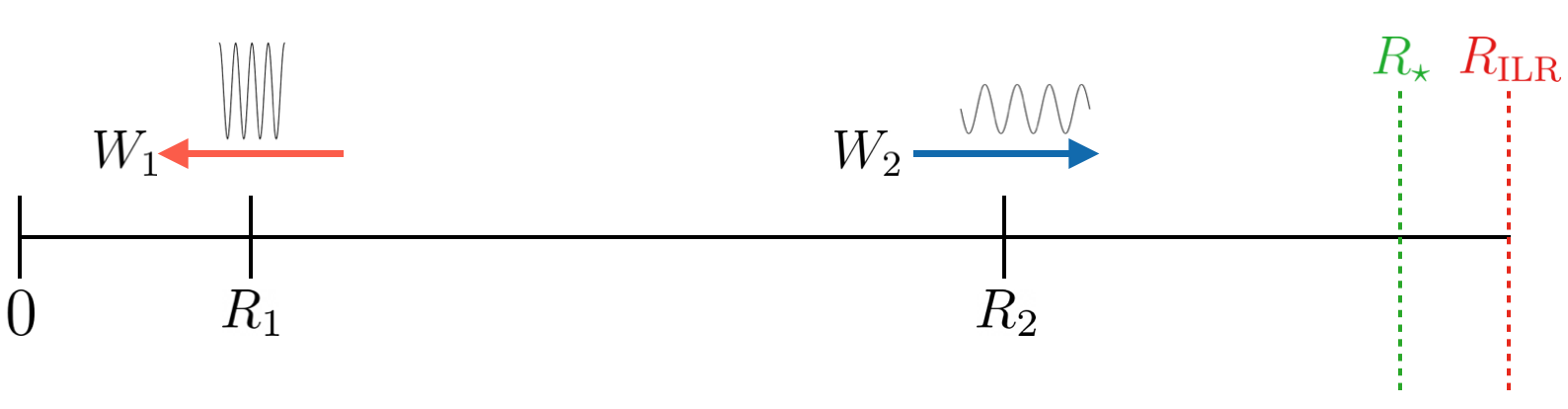}
    \caption{Schematic diagram of excitation of waves in the region $[R_1,R_2]$. $W_1$ and $W_2$ are the waves excited by the barred potential in this region. Arrows indicate the direction of propagation. See Sect.~\ref{sec:excitation} for more details.}
    \label{fig:bc}
\end{figure}

We numerically solve Eq.~\eqref{eq:mainode2} in a few selected cases. The goal is to calculate the amplitude of density waves excited by the bar potential in a region $[R_1, R_2]$ where $R_1<R_2<R_\star$, to illustrate how the amplitude depends on $\cs$ and on the unperturbed density profile $\rho_0$. A sharp edge might be present inside $[R_1, R_2]$.  The appropriate boundary conditions are ``radiation'' boundary conditions (see Fig.~\ref{fig:bc}). Causality requires that waves propagate away from the region $[R_1, R_2]$, because a solution in which waves come towards it would require a source of waves outside this region. Therefore, the correct solution to our problem contains waves propagating inwards at $R=R_1$, and outwards at $R=R_2$. These are schematically shown as the two waves $W_1$ and $W_2$ in Fig.~\ref{fig:bc}. The goal is to calculate the amplitude of $W_1$ and $W_2$.

Although the numerical solutions of Eq.~\eqref{eq:mainode2} described in these section are exact, to impose the boundary conditions we need to use the results of the WKB analysis as we need to identify the direction of propagation of the waves. We proceed as follows. Let $g_1$ be an exact solution of Eq.~\eqref{eq:mainode2} that satisfies the radiation boundary condition. We assume that at $R_1$ and $R_2$ the conditions \eqref{eq:eps1} and \eqref{eq:condgQ} are valid, so in a neighbourhood of these points we can decompose the solution as the sum of the WKB solution \eqref{eq:WKB1} and of the ``equilibrium'' solution \eqref{eq:gQ}. Therefore in a neighbourhood of $R_1$ we can write
\begin{equation} \label{eq:IC}
    g_1(R) \simeq  g_Q + W_1 \,,
\end{equation}
and in a neighbourhood of $R_2$
\begin{equation}
    g_1(R) \simeq g_Q + W_2 \,,
\end{equation}
where $W_1$ and $W_2$ are trailing and leading WKB waves respectively (Eq.~\ref{eq:WKB1}): 
\begin{align}
 W_1(R)  & = \frac{C_1}{\sqrt{K(R)}} \exp \left[i\int_{R_0}^R K(s) \di s \right] \,, \\
 W_2(R) & = \frac{C_2}{\sqrt{K(R)}} \exp \left[-i\int_{R_0}^R K(s) \di s \right]\,.
 \end{align}
To find $C_1$ and $C_2$, we use the shooting method. We start from $R=R_1$ with an initial guess for $C_1$ (which is a complex number, so the guess involves two real numbers) and initial conditions given by Eq~\eqref{eq:IC}, and integrate until $R_2$. At $R_2$ we decompose $g_1-g_Q$ into its WKB components. This decomposition is unique and can be found by equating $g_1-g_Q$ and $\di (g_1-g_Q)/\di R$ with Eq.~\eqref{eq:WKB1} and its derivative and solving the resulting algebraic system of two equations in the two unknowns that give the amplitude of the two waves. We then vary the initial guess for $C_1$ until the solution at $R_2$ only contains an outgoing wave. The amplitude of the latter gives $C_2$.

\begin{figure*}
	\includegraphics[width=\columnwidth]{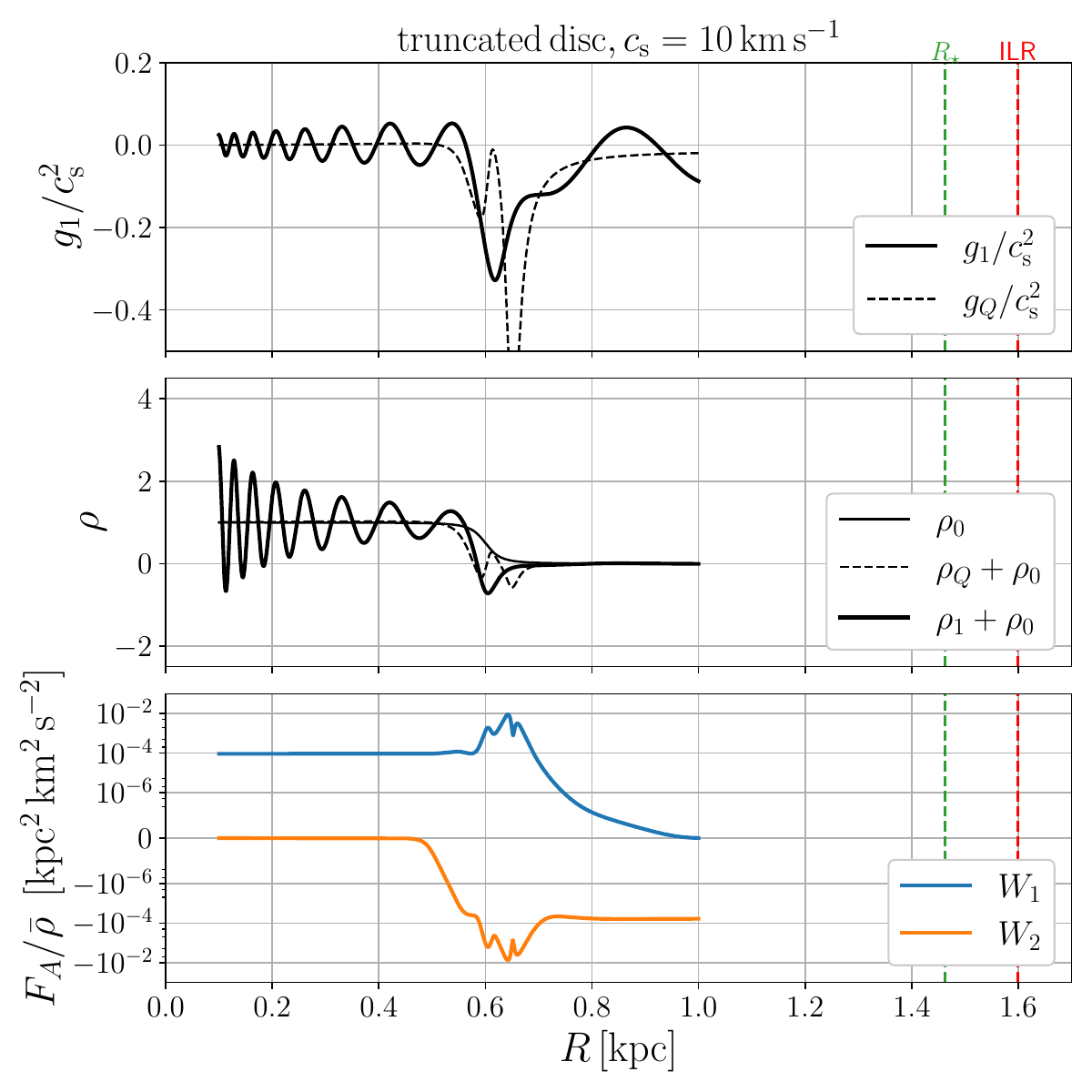}\includegraphics[width=\columnwidth]{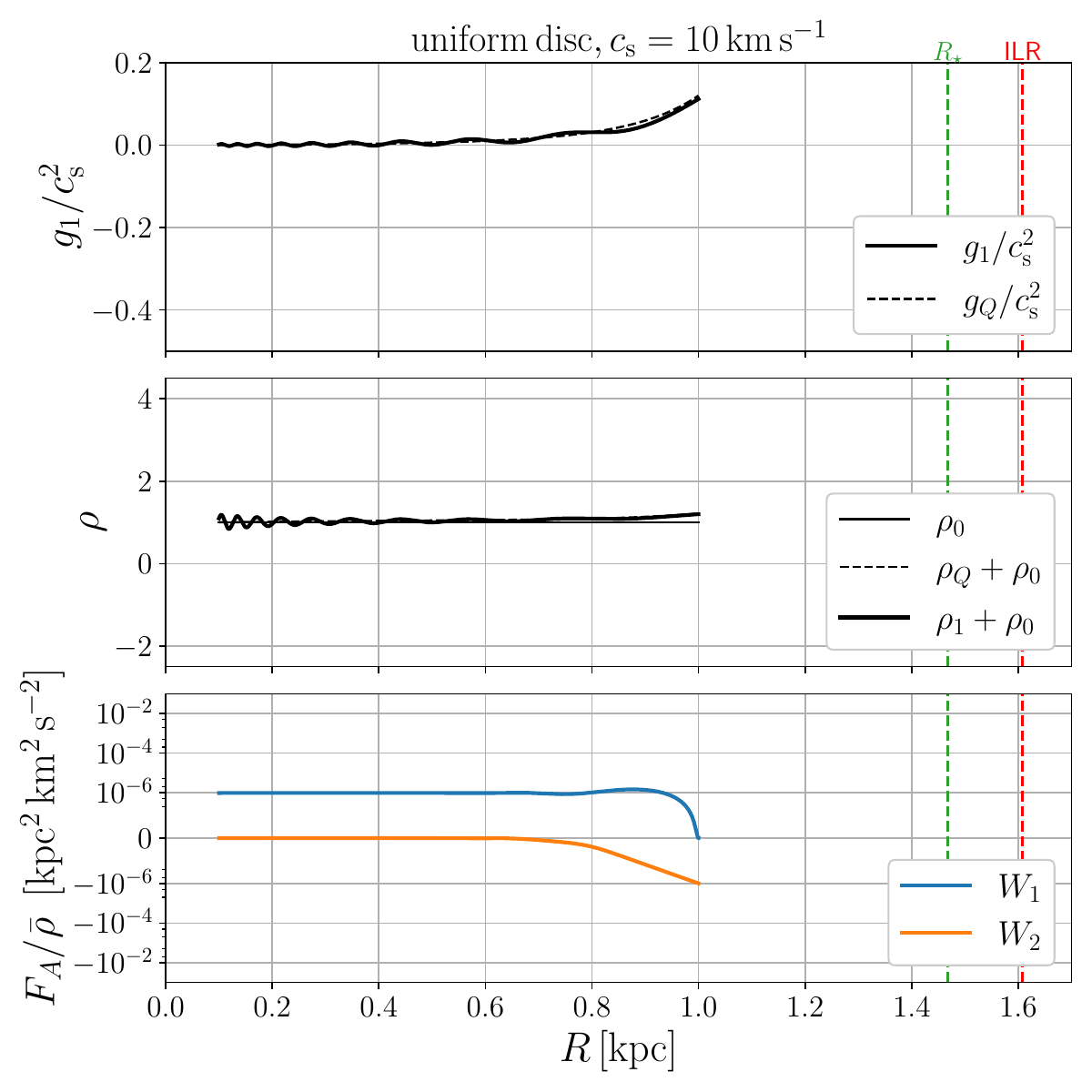}
 	\includegraphics[width=\columnwidth]{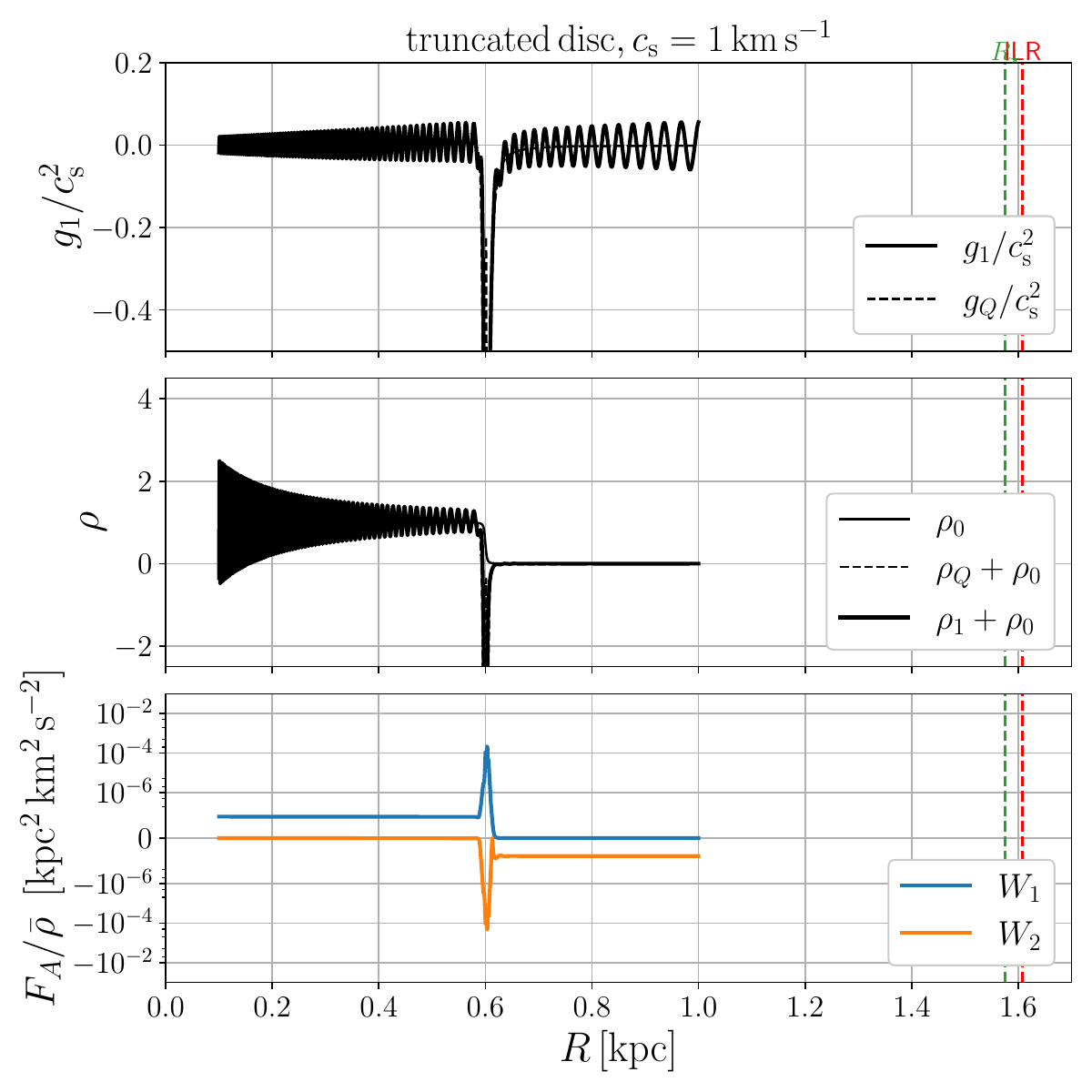}\includegraphics[width=\columnwidth]{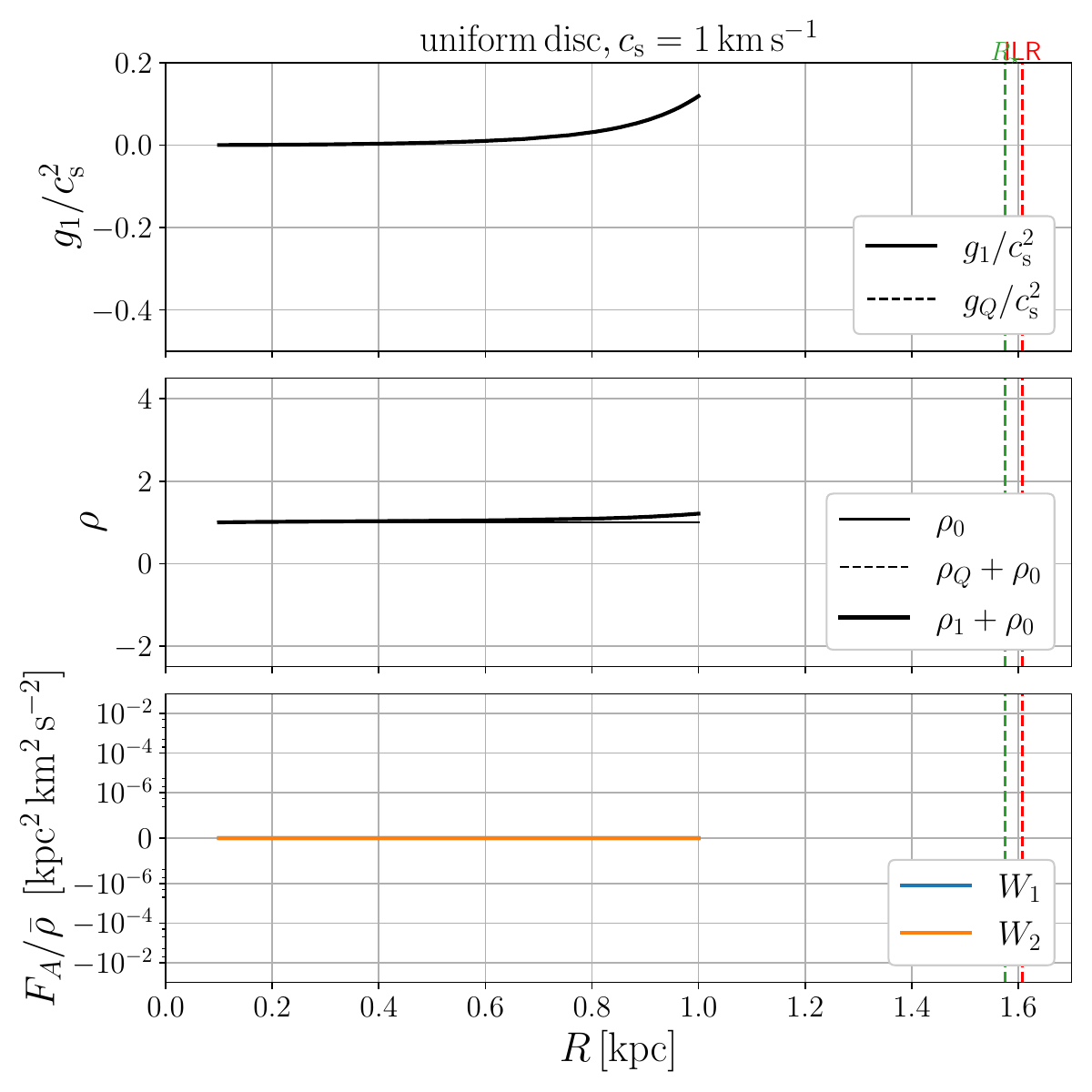}
    \caption{Waves excited by the bar potential in the region $[R_1,R_2]=[0.1,1.0]\kpc$ in uniform and truncated discs in the linear approximation. Waves excited in uniform discs are small, while waves excited at the sharp edge of a truncated disc are much larger. For each of the four cases shown, the three panels from top to bottom display the following. \emph{Top}: the solution of Eq.~\eqref{eq:mainode2} with radiation boundary conditions at $R=0.1\kpc$ and $R=1.0\kpc$ (full black line), and the approximate ``instantaneous equilibrium'' solution $g_Q$ given by Eq.~\eqref{eq:gQ} (dashed black line). The oscillations of $g_1$ around $g_Q$ give the amplitude of the excited waves. \emph{Middle}: the thin full black line shows the unperturbed density profile $\rho_0$, which can be either a uniform disc ($\rho_0=1$) or a truncated disc given by Eq.~\eqref{eq:edge} with $R_{\rm edge}=0.6\kpc$ and $\Delta R=0.03\kpc$ (for $\cs=10\kms$) or $\Delta R=0.003\kpc$ (for $\cs=1\kms$). The dashed and full thick black lines show the total density (unperturbed + perturbation) that corresponds to the solutions $g_1$ and $g_Q$ shown in the top panel. \emph{Bottom}: the flux of angular momentum associated to the two WKB waves into which $g_1-g_Q$ can be decomposed. See Section~\ref{sec:excitation} for more details.}
    \label{fig:integrate_example}
\end{figure*}

Figure~\ref{fig:integrate_example} shows the result of this procedure applied to $[R_1,R_2]=[0.1,1.0]$ for a uniform (left) and truncated disc profile (right), and for two different values of the sound speed. The truncated disc profile is chosen so that the width of the edge is comparable to the wavelength ($\lambda=2\pi/K$) at the edge. The edge cannot be much smaller than this without violating the Rayleigh criterion (Sect.~\ref{sec:unperturbed}). The top panel shows the solution $g_1$, the middle panel the corresponding density profiles, and the bottom panel the flux of angular momentum associated with the waves $W_1$ and $W_2$ obtained performing the WKB decomposition as a function of $R$.
The amplitude of the excited waves is given by the oscillations around the equilibrium solution $g_Q$ (dashed line). 

The figure illustrates the following points:
\begin{itemize}
    \item  Waves excited in uniform discs are weak (right panels). The amplitude is essentially zero at $\cs=1\kms$, while it is small but visible at $\cs=10\kms$. This reflects the fact that the approximate solution $g_Q$ is very accurate at low sound speed, but is less accurate when the sound speed is slightly larger. Physically, the reason why stronger waves are excited for larger $\cs$ is that the wavelength $\lambda$ of WKB increases with $\cs$ (Fig.~\ref{fig:Kcs}), so that waves couple more effectively to the forcing term $Q$ which varies on large scales. As we shall see in Sect.~\ref{sec:formation}, waves excited in uniform discs are too weak to remove the angular momentum necessary to open the gap.    
    \item Waves excited at the edge of a truncated disc are strong (left panels). The total density becomes negative near the edge ($\rho_1+\rho_0<0$), indicating that the linear approximation breaks down. The waves become highly non-linear and in reality they will develop shocks very quickly near the edge, as indeed seen in the simulations of Sect.~\ref{sec:numericalexperiments}. The amplitude of the waves is similar at $\cs=1\kms$ and $\cs=10\kms$, but the flux of angular momentum is much larger for $\cs=10\kms$. This will be explained by Eq.~\eqref{eq:FAimp} below.
\end{itemize}
There is a simple explanation for why strong waves are excited at sharp edges. When the background density $\rho_0(R)$ varies rapidly, such as at the edge of a disc, the forcing term $Q(R)$ on the right-hand side of Eq.~\eqref{eq:mainode2} will have a localised bump on the same scale (see dashed line in the middle panel of Fig.~\ref{fig:coefficients}). This localised bump acts like an impulsive force. In the analogy with the harmonic oscillator, this force will impart a finite amount of ``momentum'' equal to the integral of the force. The amplitude of the resulting oscillations gives the amplitude of the waves excited at the edge.

\subsubsection{Analytical estimate of the waves excited at the edge of a truncated disc} \label{sec:analyticedge}
In this section, we derive an analytical estimate for the amplitude of waves excited at a sharp edges.

Consider an edge at $R_{\rm edge}$ of width $R_{\rm out} - R_{\rm in} = \Delta R$, where $R_{\rm in}$ and $R_{\rm out}>R_{\rm in}$ are the two extremities of the region over which the edge extends (see Fig.~\ref{fig:schematic}). The shape of the edge can be arbitrary. The edge is assumed to be thin but not too thin, otherwise the unperturbed density profile would violate the Rayleigh criterion \eqref{eq:Rayleigh} and become unstable. In practice, considering the Lin-Shu approximation (Eq.~\ref{eq:KLS}) this means that the edge should not be thinner than approximately one wavelength, $\lambda = 2\pi/K$.

Away from the edge and from turning and singular points, , i.e.\ at $R<R_{\rm in}$ and $R_{\rm out}<R<R_\star$, the general solution of Eq.~\eqref{eq:mainode2} can be approximated as the sum of the WKB solution \eqref{eq:WKB1} and of the particular solution \eqref{eq:gQ}. We assume that waves are excited only near the edge, i.e. at $R_{\rm in}<R<R_{\rm out}$. We impose radiation boundary conditions, so that at $R<R_{\rm in}$ we have only the trailing wave and at $R>R_{\rm out}$ only the leading wave. Therefore outside the edge we write:
\begin{equation}
\label{eq:g1WKBinout}
g_1(R) = 
\begin{cases}
\frac{C_{\rm in}}{\sqrt{K(R)}} \exp \left[i\int_{R_0}^R K(s) \di s \right] + \frac{Q}{K^2}, & {\rm for\;} R<R_{\rm in} \,,\\
\frac{C_{\rm out}}{\sqrt{K(R)}} \exp \left[-i\int_{R_0}^R K(s) \di s \right] + \frac{Q}{K^2}, & {\rm for\;} R>R_{\rm out}\,.
\end{cases}
\end{equation}
The constants $C_{\rm in}$ and $C_{\rm out}$ will be determined by solving Eq.~\eqref{eq:mainode2} near the edge and matching the two solutions.

Near the edge, the forcing term $Q$ varies rapidly, violating condition \eqref{eq:condgQ}, and the equilibrium solution \eqref{eq:gQ} fails (dashed line in the middle panel of Fig.~\ref{fig:coefficients}). To solve Eq.~\eqref{eq:mainode2} near the edge, we proceed as follows. We assume that $K$ is approximately constant across the edge, i.e.\ in the range $R_{\rm in}<R<R_{\rm out}$. This is justified by our assumption that the edge is relatively sharp (see also the black dashed line in the top panel of Fig.~\ref{fig:coefficients}). Under this assumption, Eq.~\eqref{eq:mainode2} can be solved using the method of variation of parameters. The general solution is:
\begin{align}
g_1(R) & = A_1 e^{iKR} + A_2 e^{-iKR} + \nonumber\\
\label{eq:g1neargen}
& - \frac{ie^{iKR}}{2K}\int_{R_0}^{R}Q(s)e^{-iKs}\di s + \frac{ie^{-iKR}}{2K}\int_{R_0}^{R}Q(s)e^{iKs}\di s \;,
\end{align}
where $K\simeq K(R_{\rm in})\simeq K(R_{\rm out})$. The constants $A_1$ and $A_2$ are determined by the condition that the solution contains only waves travelling inwards at radii $R<R_{\rm in}$, and waves travelling outwards at radii $R_{\rm out}<R$. These calculations are reported in Appendix~\ref{appendix:calc1}. We find:
\begin{align}
\label{eq:A1}
A_1 & = \frac{Q_{\rm out}}{2K^2}e^{-iKR_{\rm out}} + \frac{i}{2K}\int_{R_0}^{R_{\rm out}}Q(s)e^{-iKs}\di s \\
\label{eq:A2}
A_2 & = \frac{Q_{\rm in}}{2K^2}e^{iKR_{\rm in}} - \frac{i}{2K}\int_{R_0}^{R_{\rm in}}Q(s)e^{iKs}\di s \;,
\end{align}
where $Q_{\rm out}=Q(R_{\rm out})$ and $Q_{\rm in}=Q(R_{\rm in})$.

Both Eq.~\eqref{eq:g1WKBinout} and Eq.~\eqref{eq:g1neargen} are valid solutions of Eq.~\eqref{eq:mainode2} in a neighbourhood of $R_{\rm in}$ and in a neighbourhood of $R_{\rm out}$ (shaded regions in Fig.~\ref{fig:schematic}). Matching these two solutions gives (see Appendix ~\ref{appendix:calc1}):
\begin{align}
C_{\rm in} = -C_{\rm out}^* & = \frac{ie^{iKR_0}}{2K^{1/2}} \int_{R_{\rm in}}^{R_{\rm out}}Q(s)e^{-iKs}\di s + \nonumber\\
& + \frac{Q_{\rm out}}{2K^{3/2}}e^{iK\left(R_0-R_{\rm out}\right)} - \frac{Q_{\rm in}}{2K^{3/2}}e^{iK\left(R_0-R_{\rm in}\right)} \;.
\label{eq:Cinout}
\end{align}
The coefficients $C_{\rm in}$ and $C_{\rm out}$ give the amplitude of density waves excited at the edge. It can be shown that the absolute values $|C_{\rm in}|$ and $|C_{\rm out}|$ are independent of $R_0$, as they should since the angular momentum flux at the edge should be independent of this arbitrary radius.

The rapid variation of $Q$ near the edge is what generates density waves. The coupling between the forcing term $Q$ and the density waves is expected to be maximum when the scale-length over which $Q$ varies is comparable to the wavelength of the waves $\lambda=2\pi/K$ (similarly to the toy problem in Appendix~\ref{appendix:toymodel}), i.e.\ when $\Delta R \simeq \lambda$.

We can use Eqs.~\eqref{eq:FAWKB2} and \eqref{eq:Cinout} to calculate the angular momentum flux carried by the waves excited at the edge. To obtain a closed formula it is necessary to make some further assumptions on the edge. If the edge of the disc is marginally stable to the Rayleigh criterion \eqref{eq:Rayleigh}, one has $|R_{\rm out}-R_{\rm in}|\sim \cs/\Omega \sim 1/K$. This is the sharpest edge that can be constructed without making the unperturbed density distribution unstable. Then the exponential $\exp(- iKs)$ in the integral of Eq.~\eqref{eq:Cinout} is nearly constant. As shown in Appendix \ref{sec:impulse}, in this case Eq.~\eqref{eq:Cinout} reduces to:
\begin{equation}
\label{eq:Capprox}
\left|C_{\rm in}\right| \simeq \left|C_{\rm out}\right| \simeq \left[\left(\frac{R\rho_0}{K\left|D\right|}\right)^{1/2}\left|\frac{\di\Phi_1}{\di R} +\frac{2\Omega}{\Omega-\Omega_{\rm p}}\frac{\Phi_1}{R}\right|\right]_{R=R_{\rm edge}} \;.
\end{equation}
Note that this is essentially the impulse approximation, i.e.\ we have assumed that the force $Q$ gives an instantaneous ``kick'' at $R=R_{\rm edge}$. Using Eq.~\eqref{eq:FAWKB}, the flux of angular momentum of waves excited at a sharp edge is then
\begin{equation}
\label{eq:FAimp}
F_{\rm A} \simeq m\pi \left[\left(\frac{R\rho_0}{K\left|D\right|}\right)\left(\frac{\di\Phi_1}{\di R} +\frac{2\Omega}{\Omega-\Omega_{\rm p}}\frac{\Phi_1}{R}\right)^2\right]_{R=R_{\rm edge}} \;.
\end{equation}
Eq.~\eqref{eq:FAimp} is correct when the distance of the edge from the inner Lindblad resonance is larger than approximately one wavelength, i.e. $K|R_{\rm edge}-R_{\rm ILR}|\gg 1$. At $|R_{\rm edge}-R_{\rm ILR}| = \lambda/(2 \pi^2)=1/(\pi K)$ and approximating $D(R)\simeq (R-R_{\rm ILR})(\di D/\di R)$, as appropriate near the ILR where $D$ vanishes, Eq.~\eqref{eq:FAimp} becomes identical to Eq.~(46) of \cite{Goldreich1979} which gives the flux of angular momentum of waves excited at the resonance.

\section{The formation of nuclear rings} \label{sec:formation}

We are now ready to put everything together and describe our picture of the formation of nuclear rings. For simplicity, we illustrate our scenario by starting from a uniform density distribution that extends from $R=0$ to $R\gg R_{\rm ILR}$. This is essentially the same situation as in simulation 04\_Large shown (Fig.~\ref{fig:sigma} and Sect.~\ref{sec:numericalresults2}). The formation of the ring can be schematically divided into two stages, which depend on the distance of the edge of the gas disc from the ILR. The regions corresponding to the two stages are marked in Fig.~\ref{fig:schematic}. The simulations 01-05 shown in Fig.~\ref{fig:manyrho} only include the second stage.

\subsection{First stage ($|R_{\rm edge}-R_{\rm ILR}|\lesssim \lambda)$} \label{sec:firststage}

In the first stage, a trailing spiral wave is excited near the ILR by the external bar potential. This is the regime analysed by \citet{Goldreich1979}. The wave travels inwards but for realistic strengths of the bar potential it very quickly becomes non-linear and develops into a shock. The wave then dissipates, depositing its (negative) angular momentum into the gas disc (i.e., removing angular momentum from the gas disc).\footnote{Linear waves of small amplitude travel to the centre without affecting the unperturbed density of the disc. It is only when they become non-linear that they can dump their angular momentum in the unperturbed disc.} This reduces the angular momentum of the disc, causing the gas to move inward. A gap opens around the ILR.

The width of the gap opened in the first stage is the range of validity of the calculations of \citet{Goldreich1979}, which is approximately one wavelength, i.e.\ $|R_{\rm edge}-R_{\rm ILR}|\sim \lambda$, where $\lambda=2\pi/K$. Using the Lin-Shu approximation  (Eq.~\ref{eq:KLS}) and approximating $D\simeq (R-R_{\rm ILR})(\di D/\di R)$ (recall that $D=0$ at the resonance) we have $\lambda \sim \cs / |D|^{1/2} \sim \cs/|(R_{\rm edge}-R_{\rm ILR})(\di D/\di R)|^{1/2}$ and therefore $|R_{\rm edge}-R_{\rm ILR}|\sim |\cs^2/(\di D/\di R)|^{1/3}\sim (\cs/v_0)^{2/3}R_{\rm ILR}$. The size of the gap is therefore much smaller than the radius of the resonance, and increases for increasing sound speed.

The velocity at which the edge of the gap moves can be estimated by dividing the flux of angular momentum of the waves, $F_{\rm A}$, by the amount of angular momentum per unit radius in the unperturbed disc, $2\pi \rho_0 R^3\Omega$: 
\begin{equation}
\label{eq:evogen}
\frac{\di R_{\rm edge}}{\di t} =-\left[\frac{F_{\rm A}}{2\pi \rho_0 R^3\Omega}\right]_{R=R_{\rm edge}} \;.
\end{equation}
The angular momentum flux $F_{\rm A}$ during the first stage can be estimated using Eq.~(46) of \citet{Goldreich1979}, bearing in mind that these calculations are valid in the linear approximation and should not be expected to be too accurate for the highly non-linear waves excited by a strong bar potential considered here. Taking into account that $m=2$, we find
\begin{equation}
\label{eq:evonear}
\boxed{
\frac{\di R_{\rm edge}}{\di t} = -\left[\left(\frac{\pi}{R^2\Omega (\di D/\di R)}\right)\left(\frac{\di\Phi_1}{\di R} +\frac{2\Omega}{\Omega-\Omega_{\rm p}}\frac{\Phi_1}{R}\right)^2\right]_{R=R_{\rm edge}}
}
\end{equation}
Inserting the numbers of our gravitational potential (Appendix \ref{appendix:potential}) into Eq.~\eqref{eq:evonear} we obtain $\di R_{\rm edge}/\di t\simeq 6\kms$. The duration of the first stage can be estimated by dividing the size of the gap by the velocity of the edge. Using $|R_{\rm edge}-R_{\rm ILR}|\sim (\cs/v_0)^{2/3}R_{\rm ILR}$, $v_0=220\kms$, $R_{\rm ILR}=1.6\kpc$ and  the value of $\di R_{\rm edge}/\di t$ found above we obtain:
\begin{equation} \label{eq:T1}
    T_1\simeq \left(\frac{\cs}{10\kms}\right)^{2/3} 30 \Myr \,.
\end{equation}
This time is relatively short compared to the total timescales involved (see Fig.~\ref{fig:ringsize}). In reality, the evolution is likely to be even faster than Eq.~\eqref{eq:T1} suggests because of the non-linearity of the process. The evolution of the gap after the first stage and its final size are determined by the second stage.

\subsection{Second stage ($|R_{\rm edge}-R_{\rm ILR}|\gtrsim \lambda$)} \label{sec:secondstage}

At the beginning of the second stage there is a gap around the ILR, and the distance between the inner edge of the gap and the ILR is approximately one wavelength. Since the width of the edge at this point can be at most one wavelength (because the edge tail cannot extend beyond the ILR), the edge is ``sharp'' by definition and strong waves will be excited at its location according to the analysis in Sect.~\ref{sec:mainodeanalysis}. Similarly to the waves excited near the ILR in the first stage, the waves excited near the edge will become quickly non linear and dissipate, removing the angular momentum from the gas disc and causing the edge to move inwards. Gas will accumulate at the edge, forming a ring.

This process can be seen in action in Figs.~\ref{fig:manyrho} and \ref{fig:xcut}. The first figure shows trailing waves excited by the bar potential (see for example $t=157\Myr$). The pitch angle of these waves is in good agreement with that predicted by the WKB analysis of Sect.~\ref{sec:WKB}, indicating that these are indeed trailing waves of the same type studied in the linear analysis. Fig.~\ref{fig:xcut} shows that the amplitude of the waves decreases inwards, contrary to what would be predicted in the linear approximation (e.g. Fig.~\ref{fig:integrate_example}). This is because when the waves become strongly non-linear and develop shocks, they quickly dissipate and decrease their amplitude. This dissipation is what allows the wave to deposit their angular momentum into the gas disc.

The speed at which the edge moves during the second stage can be estimated using Eq.~\eqref{eq:evogen}, where we use Eq.~\eqref{eq:FAimp} to estimate the flux of angular momentum $F_A$ of waves excited at sharp edges. Using the Lin-Shu approximation (Eq.~\ref{eq:KLS}) to write $K\simeq |D|^{1/2}/\cs$, and $m=2$, we find:
\begin{equation}
\label{eq:evo}
\boxed{
\frac{\di R_{\rm edge}}{\di t} = -\left[\left(\frac{2\cs}{R^2\Omega\left|D\right|^{3/2}}\right)\left(\frac{\di\Phi_1}{\di R} +\frac{2\Omega}{\Omega-\Omega_{\rm p}}\frac{\Phi_1}{R}\right)^2\right]_{R=R_{\rm edge}}
}
\;.
\end{equation}
where the factor of $2$ takes into account that the outward-travelling leading wave excited at the edge will be reflected at $R=R_\star$ into an inward-travelling trailing wave. Note that Eq.~\eqref{eq:evonear} and \eqref{eq:evo} only differ for the factor in the first round parentheses on the right-hand-sides, and this factor coincides in the two equations at a distance of approximately one wavelength from the ILR. This is the point where we transition from the analysis of \cite{Goldreich1979} to the analysis in Sect.~\ref{sec:analyticedge}, and from the first to the second stage.

Fig.~\ref{fig:ringsize} compares the size of the ring as a function of time predicted by Eq.~\eqref{eq:evo} to that measured in the numerical experiments of Sect.~\ref{sec:numericalresults1}. We find that the equation captures the general trends in the figure, including the fact that the edge moves faster for larger sound speed, but it tends to underestimate the speed at which it moves, especially at large sound speed. That the analytic prediction is not quantitatively accurate is not surprising considering that Eq.~\eqref{eq:evo} is derived in the linear approximation, but the waves excited at the edge are strongly non-linear (Figs.~\ref{fig:xcut} and \ref{fig:integrate_example}). The flux of angular momentum generated in the case of a uniform disc is too small to move the edge significantly over the course of several Gyr (Fig.~\ref{fig:integrate_example}).

\begin{figure}
    \includegraphics[width=\columnwidth]{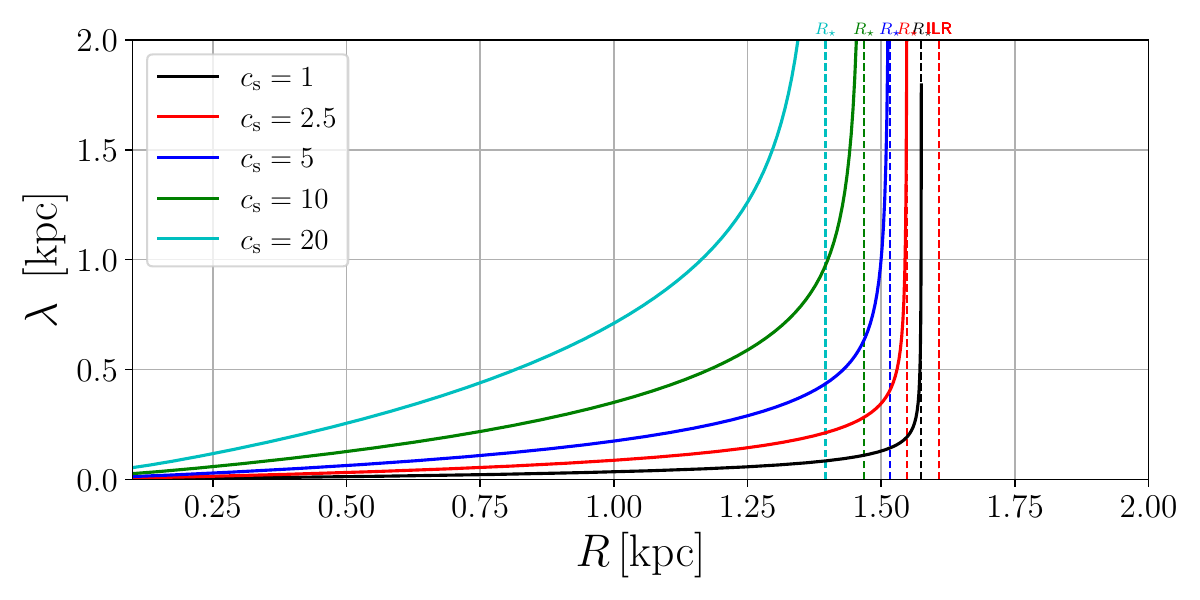}
    \caption{Wavelength of WKB waves ($\lambda = 2\pi/K$) for different values of the sound speed $\cs$. The wavenumber $K$ is given by Eq.~\eqref{eq:dispersionrelation} assuming a uniform disc $\rho_0(R)=1$.}
    \label{fig:Kcs}
\end{figure}

When does the edge stop moving? The process above continues until waves can be effectively excited at the edge, which happens when both of the following conditions are satisfied: (i) the edge is ``sharp'', i.e. the edge width is smaller than a few times the wavelength of density waves $\lambda=2\pi/K$; (ii) the gravitational potential $\Phi_1$ is sufficiently strong. The distance between the edge and the ILR poses an upper limit to the width of the edge since the edge cannot cross the ILR, $\Delta R<|R_{\rm ILR} - R_{\rm edge}|$.\footnote{Recall also that as discussed in Sect.~\ref{sec:analyticedge} the edge cannot be too thin, otherwise the system becomes Rayleigh-unstable. Thus, we expect the edge width to remain of order $\lambda$ during the shrinking process.} Therefore, when the edge is not sufficiently far from the ILR, it \emph{must} be sharp. In particular, we can expect the edge to keep moving until it is located a few wavelengths away from the ILR. Since $\lambda$ increases linearly with $\cs$ (Eq.~\ref{eq:KLS} and Fig.~\ref{fig:Kcs}), we expect the edge to move farther at larger sound speed, which explains why the ring radius depends on the sound speed.

Predicting exactly where the edge will stop, and therefore the final radius of the ring, is a difficult task. The process is highly non-linear, and the unperturbed density profile changes in a way that cannot be calculated in the linear approximation. Empirically, we find from the numerical experiments in Sect.~\ref{sec:numericalexperiments} that for our assumed gravitational potential the ring stops when one can fit approximately $7$ wavelengths $\lambda$ between $R_{\rm edge}$ and $R_{\rm ILR}$. For weaker barred potential, the edge might stop sooner if $\Phi_1$ is too small to generate sufficient flux of angular momentum at the edge.

Finally, we note that our theory satisfies all the 5 conditions that we laid out in Sect.~\ref{sec:constraints}. Conditions 1-3 are satisfied because clearly the radius of the ring depends on the rotation curve, on the non-axisymmetric part of the gravitational potential, and on the pattern speed of the bar which sets the location of the ILR. All these dependencies are also evident in Eq.~\eqref{eq:evo}. Condition 4 is satisfied because the radius of the ring depends on the sound speed of the gas in two ways: first because the speed at which the edge moves away from the ILR increases as a function of $\cs$ (see Eq.~\ref{eq:evo}), and second because the final ring size is determined by the condition that the edge should be a few wavelengths away from the ILR, which results in smaller rings at larger sound speed since as discussed above the wavelength increases with the sound speed. Condition 5 is satisfied since the excitation of density waves at the edge is a local process.

\section{Discussion} \label{sec:discussion}

\subsection{Comparison with the works of Goldreich \& Tremaine}

During the late 70's and early 80's, Peter Goldreich and Scott Tremaine published a series of papers in which they studied the dynamics of planetary rings. The calculations and physical processes studied in these works have much in common with those presented in the present paper. Here we highlight the main similarities and differences between them and the present paper.

\cite{Goldreich1978} (hereafter GT78) developed a picture for the formation of the Cassini division in Saturn's ring that has several similarities with our picture for the formation of nuclear rings described in Sect.~\ref{sec:formation}. In both cases: (i) a gap opens near the Lindblad resonance due to waves excited at the resonance; (ii) subsequent excitation of waves at the edge of the gap continues to widen the gap. The main differences are: (i) The bar potential considered here is a perturbation many orders of magnitude stronger than the one from Saturn's satellite Mimas; (ii) the sound speed is negligibly small in Saturn's problem, while the effects of finite sound speed are important in our problem (Sect.~\ref{sec:formation}); (iii) self-gravity is negligible for our case, but it is not negligible in Saturn's problem. In particular, gravity is the main means of transport of angular momentum in Saturn's problem, while advective transport through pressure is the main mechanism for transport in our problem.

\cite{Goldreich1979} (hereafter GT79) studied the excitation of density waves in a differentially rotating gas disc by a rigidly rotating external potential. Their calculations have similarities with those presented in Sect.~\ref{sec:lineardisc}, but there are three key differences: (i) GT79 assume that $\cs\to0$, while we take into account the effects of finite sound speed; (ii) GT79 assume that $\rho_0$ varies slowly; (iii) GT79 included the self-gravity of the gas disc, which we have neglected. The combination of (i) and (ii) is why GT79 find that waves can be excited \emph{only} at the resonances, while for example in Sect.~\ref{sec:excitation} we find waves excited away from the resonance. The physical explanation is the following. The external potential can couple effectively to density waves only when the wavelength of WKB waves ($\lambda=2\pi/K$) is comparable to the typical scalelength over which the forcing term $Q$ in Eq.~\eqref{eq:mainode2} varies, i.e. when $\lambda \sim Q/(\di Q / \di R)$. The quantity $Q/(\di Q / \di R)$ is determined by the external potential $\Phi_1$ and by the unperturbed density distribution $\rho_0$, and is therefore typically very large unless there are sharp edges in $\rho_0$. In the limit of vanishing sound speed, $Q/(\di Q / \di R) \gg \lambda$ everywhere except near the turning point $R_\star$ at which $\lambda \to \infty$ (see Fig.~\ref{fig:Kcs}). In the limit $\cs \to 0$ the turning point merges with the ILR (Sect.~\ref{sec:Rstar}) and $\lambda$ is small everywhere else. Thus, in this limit waves can be excited only at the resonance. For finite sound speed instead $\lambda$ can become large and comparable to $Q/(\di Q / \di R)$ away from the resonance and waves can be excited (see Fig.~\ref{fig:Kcs}). As we have seen in Sect.~\ref{sec:formation}, the effects of finite sound are important in the formation of nuclear rings.

\subsection{Relation to $x_2$ orbits}

\begin{figure}
	\includegraphics[width=\columnwidth]{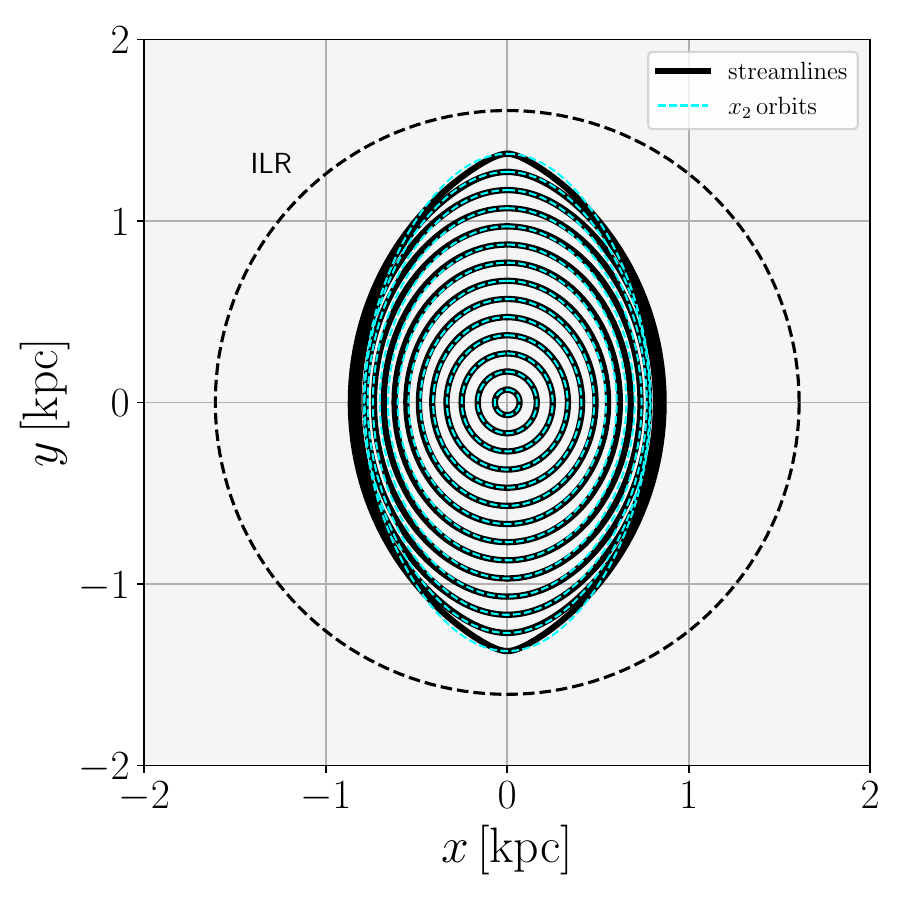}
    \caption{\emph{Black full lines}: Streamlines of the non-wave solution \eqref{eq:gQ} in the case $\cs\to 0$ and constant $\rho_0$. \emph{Cyan dashed lines}: ballistic closed $x_2$ orbits in the barred potential described in Appendix~\ref{appendix:potential}. The two are similar, showing that the density waves excited by the bar potential discussed in Sect.~\ref{sec:lineardisc} are essentially perturbations travelling on top of an $x_2$ disc.}
    \label{fig:x2}
\end{figure}

Several works have suggested a connection between nuclear ring and $x_2$ orbits \citep[e.g.\ ][]{Regan2003,Li2015,Sormani+2018b}. The $x_2$ orbits are a family of non-circular closed orbits that can exist in the central regions of a bar potential, and are elongated in the direction perpendicular to the major axis of the bar \citep[e.g.][]{ContopoulosGrosbol1989,Athan92a}. Figure~\ref{fig:x2} illustrates the relation between these orbits and the present paper. The streamlines of the ``equilibrium" solution Eq.~\eqref{eq:gQ} are very similar to closed $x_2$ orbits in the same bar potential. Therefore, the WKB waves excited by the bar potential that we studied in Sect.~\ref{sec:mainodeanalysis} travel on top of an $x_2$ gas disc. Our picture for the formation of the rings is therefore consistent with the idea that gas in nuclear rings flows on $x_2$ orbits.

\subsection{Relation with the resonant theory}

Our theory is somewhat the ``opposite'' of the resonant theory, which states that the ring forms at the ILR \citep{Combes1988,Combes1996,Buta1996}. In our theory the gas is \emph{pushed away} from the ILR rather than accumulating at it. Our theory is more consistent with the fact that the rings are typically \emph{inside} the ILR in simulations \citep[e.g.][]{Englmaier1997,Patsis2000,Kim++2012a,Sormani2015a,Li2015} and with observations that show that for example in the Milky Way the radius of the nuclear ring is $\simeq 100$-$200$pc while the ILR is at $R>500\pc$ \citep{Henshaw2022}. A key difference between our theory and all previous theories, including the resonant theory, is that we can explain the puzzling dependence of nuclear ring size on the sound speed seen in simulations.

\subsection{Brief considerations on magnetic fields and turbulent pressure} \label{sec:bfields}

The mechanism for the formation of rings described in Sect.~\ref{sec:formation} relies on waves propagated through pressure. We would therefore expect that adding magnetic fields, which create magnetic pressure in the gas, could have a similar effect as increasing the sound speed, and would therefore lead to smaller rings.

Turbulent pressure seems to have a smaller effect than ``real'' microscopic pressure on the size of nuclear rings. \cite{Salas2020} performed some numerical experiments with external turbulence driving. Their Figs. 1 and 2 show that turbulence driving changes the size of the nuclear ring by a smaller amount than an increase in the sound speed when the injected turbulent energy is comparable to the corresponding increase in thermal energy. We attribute this to the fact that, due to the presence of inelastic collisions, in a gas with turbulent pressure sound waves do not propagate as efficiently as in a gas that has the same amount of microscopic pressure.

\subsection{On the assumption of an isothermal equation of state}

Throughout this work, we have assumed an isothermal equation of state. This follows a tradition of works adopting the isothermal prescription to study the dynamics of the ISM on galactic scales and in nuclear rings \citep[e.g.][]{Roberts1969,Cowie1980,Athan92b,Englmaier1997,Fux1999,Maciejewski2004,Kim++2012a,Sormani2015a,Fragkoudi2017,Li2022}. However, the real ISM is multi-phase, turbulent and highly inhomogeneous. It is therefore natural to ask whether the isothermal prescription can capture the basic mechanism for the formation of nuclear rings.

Numerical simulations that include a multi-phase medium via thermal instabilities \citep{Sormani+2018a} as well as gas self-gravity, star formation and supernova feedback \citep{Armillotta2019,Tress2020} show that the large-scale morphology of nuclear rings is only moderately affected by the presence of this additional physics. The main differences are observed at small scales (smaller or comparable to the width of the rings), where gas condenses into molecular clouds and collapses to make star formation. The large-scale properties of the ring, such as its radius and width, can be often mimicked by using an `effective' isothermal sound speed. For example, we found that the morphology, width and radius of the nuclear ring in the simulations of \cite{Sormani+2018a}, which include a non-equilibrium chemical network that produces a two-phase medium via the thermal instability, are very similar to those obtained by replacing the non-equilibrium network and the associated cooling function with an isothermal equation of state with a low sound speed of $\cs\simeq1\kms$. This low value is because the gas in the ring is very cold in these simulations, as they did not include any sources of heating or turbulence such as stellar feedback. When star formation and stellar feedback are added to the simulations \citep[e.g.][]{Armillotta2019,Tress2020}, they heat up the gas and generate turbulent pressure, and the morphology of the rings can still be crudely mimicked by raising the isothermal sound speed (although as noted in Sect.~\ref{sec:bfields} by less than the turbulent velocity dispersion, which would be the naive way of doing it). Similarly, the effects of magnetic fields can be crudely mimicked by increasing the sound speed by summing in quadrature the typical Alfv{\'e}n speed. 

In conclusion, the isothermal prescription should be viewed as an `effective' equation of state that takes into account in a phenomenological way the additional physics via a single parameter that can be easily controlled. Ultimately, the key property that needs to be captured in this approach is the ability of the medium to propagate waves through pressure. Thus, the sound speed does not correspond to the actual kinetic temperature of the gas, but to an ‘effective’ temperature that takes into account in a crude way averaging over different phases, turbulent motions on unresolved scales, and other effects such as magnetic pressure. Reassuringly, numerical simulations suggest that the basic mechanism for the formation of the ring is well captured using this approach.

\section{Conclusion} \label{sec:conclusion}

We have used both hydrodynamical simulations and analytical calculations of linear disc dynamics to construct a new theory for the formation of nuclear rings in barred galaxies. According to this theory, nuclear rings are an accumulation of gas at the inner edge of a gap that forms around the Inner Lindblad Resonance (ILR) of a bar potential. The gap initially opens because the bar potential excites strong trailing waves around the ILR, which remove angular momentum from the gas disc and push the gas inwards. The gap then continues to widen because the bar potential excites trailing waves at the inner edge of the gap, until the edge stops at a distance of several wavelengths from the ILR. The gas accumulates at the inner edge of the gap, forming a ring. The speed at which the gap edge moves and its final distance from the ILR, which determine the radius of the nuclear ring, depend on the gas sound speed through the dispersion relation. 

Our theory has much in common with the picture for the formation of the Cassini gap in Saturn's ring proposed by \cite{Goldreich1978}. The most important differences are that (i) the effects of finite sound speed are important in our problem, while the sound speed can be assumed to be vanishingly small in the planetary problem; (ii) we have neglected the effects of self-gravity, which are typically less important in the nuclear ring problem, but cannot be neglected in the planetary rings problem.

\section*{Acknowledgements}

MCS acknowledges financial support of the Royal Society (URF\textbackslash R1\textbackslash 221118). ES acknowledges financial support of the European Union's Horizon 2020 research and innovation program  under the Marie Sk{\l}odowska-Curie Grant agreement No.~101061217. JLS acknowledges financial support of the Royal Society (URF\textbackslash R1\textbackslash 191555).

\section*{Data Availability}

The data and code underlying this article will be shared on reasonable request to the authors.

\bibliographystyle{mnras}
\bibliography{bibliography}

\appendix

\section{External gravitational potential} \label{appendix:potential}

\begin{figure}
	\includegraphics[width=\columnwidth]{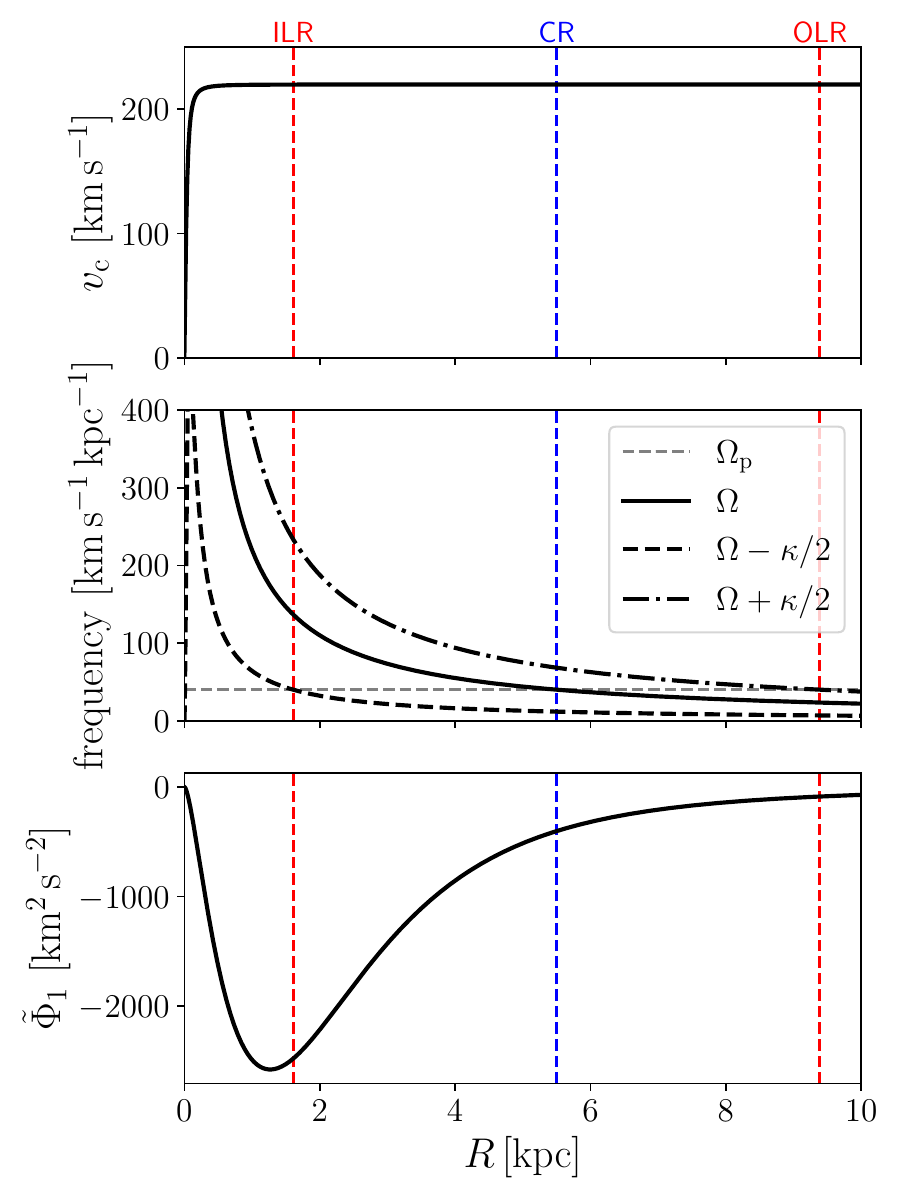}
    \caption{\emph{Top}: the circular velocity of our potential, $v_{\rm c}=(r \di \Phi_0/\di R)^{1/2}$. \emph{Middle:} the curves $\Omega$ and $\Omega \pm \kappa/2$, where $\Omega=v_{\rm c}/R$ and $\kappa$ is the epyciclic frequency (Eq.~\ref{eq:epyciclic}). The intersection of these curves with the horizontal line at $\Omega_{\rm p}$ gives the position of the resonances, indicated by vertical dashed lines. \emph{Bottom}: the quadrupole (Eq.~\ref{eq:quadrupole}).}
    \label{fig:vcirc}
\end{figure}

In this Appendix, we describe the external barred gravitational potential that is used throughout the paper. Consider a rigidly rotating potential of the following form
\begin{equation} \label{eq:potential}
    \Phi(R,\theta) = \Phi_0(R) + \tilde{\Phi}_1(R) \cos(2\theta) \,,
\end{equation}
where $(R,\theta)$ are standard polar coordinates. This represents the simplest possible barred potential, consisting of a monopole and a quadrupole. For the monopole, we take a simple axisymmetric logarithmic potential
\begin{equation} \label{eq:monopole}
    \Phi_0(R) = \frac{v_0^2}{2} \log\left(R^2 + R_{\rm c}^2\right)
\end{equation}
where $v_0=220\kms$ and $R_{\rm c}=0.05\kpc$. The logarithmic potential is convenient because the rotation curve is rising at small $R$ and is flat at $R\gg R_{\rm c}$, roughly consistent with the rotation curves observed in many disk galaxies \citep[e.g.][]{Lang2020}. For the quadrupole, we employ the analytic density-potential pair described in Appendix A of \cite{Sormani+2018b}:
\begin{equation} \label{eq:quadrupole}
    \tilde{\Phi}_1(R) = - A (v_0 e)^2 f\left(\frac{R}{R_q}\right) 
\end{equation}
where $A=0.4$ is a dimensionless parameter that quantifies the bar strength, $e=2.71[\dots]$ is Euler's number, $v_0=220\kms$ is the same as in Eq.~\eqref{eq:monopole}, $R_q=1.5\kpc$ is the radial scalelength, and $f$ is the following function:
\begin{equation} \label{eq:F}
    f(x) = \frac{3-e^{-2x}\left(2x^4 + 4x^3 + 6x^2 + 6x+3\right) + 4 x^5 \mathrm{E}_1(2x)}{20x^3} \,,
\end{equation}
where $\mathrm{E}_1(x)$ is the exponential integral function, a special function defined as
\begin{equation}
    \mathrm{E}_1(x) = \int_x^\infty \frac{e^{-t}}{t}\,\di t \,.
\end{equation}
This quadrupole reproduces well those generated by $N$-body exponential bars. We assume that the potential rotates with pattern speed $\Omega_{\rm p}=40 \kms \kpc^{-1}$. This places the inner Lindblad resonance at $R_{\rm ILR}=1.61 \, \kpc$, the corotation resonance at $R=5.5\kpc$, and the outer Lindblad resonance at $R_{\rm OLR}=9.39\, \kpc$. Figure~\ref{fig:vcirc} shows the circular velocity (top), the resonance diagram (middle) and the quadrupole (bottom) of our potential.

\section{Excitation of 1D waves by an oscillating Gaussian potential (toy model)} \label{appendix:toymodel}

In this Appendix, we describe a toy model that shares various similarities with the actual problem studied in the main text but has the advantage that it can be solved fully analytically. This toy model is helpful for understanding why the amplitude of the waves excited by an external potential can depend very strongly on the gas sound speed.

We can draw the following correspondences between this toy problem and the problem studied in the main text (i) 1D sound waves correspond to spiral density waves, and in particular to the WKB waves discussed in Sect.~\ref{sec:WKB} (ii) The 1D Gaussian potential corresponds to the bar potential; (iii) The linear momentum of plane waves plays a similar role to the angular momentum of the spiral density waves; (iv) Eq.~\eqref{eq:exc_07} is the analog of Eq.~\eqref{eq:mainode2}.

\subsection{Statement of the problem}

Consider a 1D isothermal fluid at rest with uniform density $\rho_0$. Our goal is to study the waves excited in this medium by a ``small'' time-varying external potential $\Phi(x,t)$.

The equations of motion of this system are the same as Eqs.~\eqref{eq:continuity}, \eqref{eq:euler} and \eqref{eq:isothermaleos} where the gradient is replaced by $\di/\di x$ since the problem is one dimensional. We linearise these equations around the background state by writing $\rho(x,t)=\rho_0+\rho_1(x,t)$ and $v(x,t)=v_1(x,t)$ and keeping only the first-order terms in the quantities with subscript 1. We obtain:
\begin{align}
& \pa_t \rho_1 + \rho_0 ( \pa_x v_1)  = 0 \,, \label{eq:exc_01} \\
& \pa_t v_1  = -\cs^2 \frac{\pa_x \rho_1}{\rho_0} - \pa_x \Phi \,.\label{eq:exc_02}
\end{align} 
Without loss of generality, we can write all variables as
\begin{align}
F(x,t)	& = \tilde{F}(x) \exp(- i \omega t) \,. \label{eq:exc_03}
\end{align}
Where $\tilde{F}$ are complex quantities. We use complex notation for mathematical convenience, but it is understood that the physical quantities are given by the real part. Substituting all perturbation variables in the form \eqref{eq:exc_03} into \eqref{eq:exc_01} and \eqref{eq:exc_02}, and omitting the symbol $\,\tilde{}\,$ hereafter for simplicity of notation, we obtain
\begin{align}
& - i \omega \rho_1 + \rho_0 ( \pa_x v_1)  = 0 \label{eq:exc_04} \\
& - i \omega v_1 = - \cs^2 \frac{\pa_x \rho_1}{\rho_0} - \pa_x \Phi(x) \label{eq:exc_05}
\end{align} 
Isolating $v_1$ from \eqref{eq:exc_05} and introducing the variable $s_1=\rho_1/\rho_0$ we have
\begin{equation} \label{eq:exc_06}
v_1 = \frac{ \cs^2  (\pa_x s_1) + (\pa_x \Phi)}{i \omega} 
\end{equation}
Substituting \eqref{eq:exc_06} into \eqref{eq:exc_04} we obtain an ODE for $s_1$:
\begin{equation} \label{eq:exc_07} \boxed{
\omega^2 s_1 + \cs^2 \frac{\di^2}{\di x^2}s_1  = F}
\end{equation}
where
\begin{equation}
F(x) = - \pa_x^2 \Phi\,.
\end{equation}
Eq.~\eqref{eq:exc_07} is the equation of a forced harmonic oscillator. Now consider an oscillating Gaussian potential of the form:
\begin{align} \label{eq:gaussphi}
\Phi(x,t) & = \Phi_1 \exp\left[ -\left(\frac{x}{x_0}\right)^2\right]  \exp(-i \omega_0 t) 
\end{align}
where $\Phi_1$ is the strength of the potential, $x_0$ is the width of the Gaussian perturbation, $\omega_0$ is the oscillation frequency. Since in the linear approximation there is no coupling between modes at different frequencies, only modes with frequency $\omega=\omega_0$ will be excited by this potential. Hence we assume $\omega=\omega_0$ hereafter. Introducing the dimensionless coordinate $\xi=x/x_0$ and using \eqref{eq:gaussphi}, Eq.~\eqref{eq:exc_07} becomes:
\begin{equation} \label{eq:exc_100}
\boxed{
 \frac{\di^2 s_1}{\di \xi^2} + a^2 s_1 =  b a^2 K(\xi) \,}
\end{equation}
where 
\begin{equation}
K(\xi) = (1 - 2 \xi^2) \exp\left[-\xi^2\right] \,,
\end{equation}
and we have introduced the following dimensionless parameters:
\begin{align}
a & = \frac{\omega_0 x_0}{\cs} \, \label{eq:exc_100bis}, \\
b & = \frac{2\Phi_1}{\omega_0^2 x_0^2} \,.
\end{align}
 The parameter $a$ is the inverse of the sound speed, normalised with the typical scale-length and frequency of the problem. The parameter $b$ is the normalised strength of the external potential.
 
\subsection{Analytical solution}
The general solution of Eq.~\eqref{eq:exc_100} is
\begin{align}
s_1(\xi)  & = C_1 \exp(i a \xi) + C_2 \exp(-ia \xi) +  b W(\xi,a) \,, \label{eq:exc_101}
\end{align}
where $C_1$ and $C_2$ are arbitrary constants and
\begin{align}
W(\xi,a) & = -\frac{1}{2} a^2 e^{-\xi^2} + X(\xi,a) \,, \label{eq:exc_102} \\
X(\xi,a) & = - i \alpha \left[ e^{i a \xi} \, \operatorname{erf}\left(\xi + i \frac{a}{2} \right) - e^{- i a \xi} \, \operatorname{erf}\left( \xi  - i \frac{a}{2}\right) \right] \,, \\
\alpha    & =  \frac{\sqrt{\pi}}{8} a^3 e^{-a^2/4}\,.
\end{align}
Here, $\operatorname{erf}(z) = \frac{2}{\sqrt{\pi}} \int_0^z e^{-t^2} \, \di t$ is the error function, which is defined for complex argument $z$ (to evaluate the integral, you can choose any integration path in the complex plane that leads to $z$). Note however that the functions $X$ and $W$ are real because the erf function has the properties $\operatorname{erf}(\overline{z}) = \overline{ \operatorname{erf}(z)}$ and $\operatorname{erf}(-z) = \operatorname{erf}(z)$, where the bar denotes the complex conjugate.

\begin{figure}
	\includegraphics[width=\columnwidth]{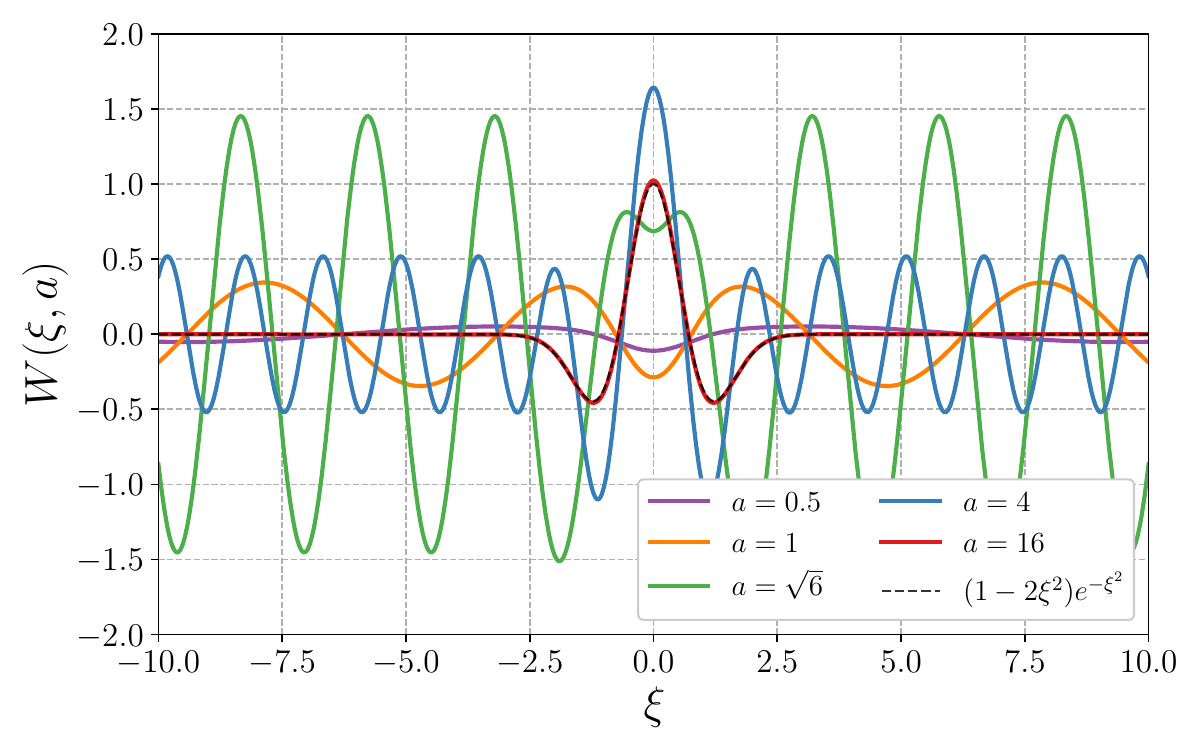}
    \caption{The function $W(\xi,a)$ defined by Equation~\eqref{eq:exc_102} for various values of $a$.}
    \label{fig:toy}
\end{figure}

Figure \ref{fig:toy} plots the function $W(\xi,a)$ for various values of $a$. In the limit $\xi\to \pm \infty$ we have that $\operatorname{erf}(\xi + i c/2) \to \pm 1$ for any fixed $c$, so $W(\xi,a) \to \mp i \alpha [ e^{i a \xi}- e^{- i a \xi}  ] $.  Therefore $W(\xi,a)$ becomes a plane wave when $\xi\to\pm\infty$ (as one would expect). In the limit $a\to\infty$ the $W$ tends to the forcing term $K$ on the right-hand-side of Eq.~\ref{fig:toy}.

What is the amplitude of the waves that are excited by the external potential \eqref{eq:gaussphi}? In order to answer this question we have to determine the constants $C_1$ and $C_2$ in Eq.~\eqref{eq:exc_101} by imposing appropriate boundary conditions. Causality requires that for large $|x|$ the waves propagate ``away'' from the potential (this is the same boundary condition that is used to derive retarded potential in electrodynamics).  A solution in which the waves come from infinity towards the potential would instead require a source at infinity, which is unphysical. Therefore, we impose that the solution propagates towards positive $\xi$ as $\xi\to+\infty$ and towards negative $\xi$ as $\xi\to-\infty$. To see in which direction the solution \eqref{eq:exc_101} is travelling, we look at its time-dependence by reattaching the factor $\exp(- i \omega_0 t)$ to it:
\begin{align}
s_1(\xi,t)  & = \left[ C_1 e^{i a \xi} + C_2 e^{-ia \xi} + b W(\xi,a) \right]  e^{- i \omega_0 t} \,. \label{eq:exc_103} 
\end{align}

A plane wave of the form $e^{i a \xi - i \omega_0 t}$ ($e^{-i a \xi - i \omega_0 t}$) travels towards positive (negative) $\xi$. The solution that satisfies our ``radiation'' boundary conditions is then:
\begin{align}
s_1(\xi,t)  & = b \left[- i \alpha e^{i a \xi} - i \alpha  e^{-ia \xi} + W(\xi,a) \right]  e^{- i \omega_0 t}   \,.
\end{align}
This solution tends to $s_1(\xi,t) \to - 2 i \alpha  b e^{\pm ia \xi} e^{- i \omega_0 t}  $ for $x\to \pm \infty$. Thus, the potential excites waves with an amplitude of 
\begin{equation}
\boxed{
A = 2   b \alpha =  b \frac{\sqrt{\pi}}{4} a^3 e^{-a^2/4} .}
\end{equation}  
The key point here is that the amplitude $A$ of the excited waves has an extremely strong dependence on the sound speed $\cs \propto 1/a$. The amplitude $A$ tends to zero \emph{very} quickly both for $a\to 0$ ($\cs\to \infty$) and $a \to \infty$ ($\cs\to 0$), and (for fixed $b$) has a maximum in between at $a=\sqrt{6}$. This has a simple physical interpretation. The coupling between the external potential and sound waves in a uniform medium is strongest when the wavelength of free sound waves at the frequency of the external potential is comparable to the scale-length of the potential. This is indeed what happens, as can be seen as follows. The dispersion relation of free sound waves travelling in a uniform medium is $\omega = \cs k$, where $k=2 \pi/\lambda$ is the wavenumber and $\lambda$ is the wavelength. Therefore, the wavelength of free sound waves travelling in a uniform medium at frequency $\omega_0$ is $\lambda_0=2\pi\cs/\Omega_0$. The parameter $a=2\pi x_0/ \lambda_0$ is, apart from a numerical constant, the ratio between the scale-length of the potential and the wavelength of free sound waves at that frequency. Thus, we expect the potential to be most effective in driving waves when $a$ is of order unity. 

\section{The WKB method}
\label{appendix:WKB}

The WKB method is a method for finding approximate solutions to linear differential equations with spatially varying coefficients. Let us briefly review how it works \citep[for a more extensive review see for example][]{BenderOrszag1999}. The notation used in this Appendix is not related to the notation in the main text (for some quantities we use the same symbols). Consider the following equation:
\begin{equation} \label{eq:WKBex2_01}
\ddot{x} + \omega^2(t) x = 0\;,
\end{equation}
where $\omega(t)$ is a given function of $t$. If $\omega$ were constant, Eq.~\eqref{eq:WKBex2_01} would be the equation of a harmonic oscillator, with general solution:
\begin{equation} \label{eq:WKBsol1}
x(t) = C_1 \exp\left[ i \omega t \right] + C_2 \exp\left[- i \omega t \right]\;,
\end{equation} 
where $C_1$ and $C_2$  are arbitrary complex constants. The period of oscillation is $T=2 \pi/ \omega$. 

When $\omega(t)$ is not constant, Eq.~\eqref{eq:WKBex2_01} has in general no analytic solution. However, when $\omega(t)$ is ``slowly varying'', we can find solutions using the WKB method. By ``slowly varying'', we mean that the changes in $\omega(t)$ during an oscillation are small. This condition can be written:
\begin{equation} \label{eq:WKBcond}
\left|    \frac{\dot{\omega}}{\omega^2} \right| \ll 1 \,.
\end{equation}
Eq.~\eqref{eq:WKBex2_01} physically corresponds to a mass $m$ connected to a spring with a time-dependent spring constant $k(t)=m \omega^2(t)$. When $\omega(t)$ is varying slowly, we expect the system to instantaneously behave almost as if $\omega$ were constant, and to slowly ``morph'' over time between solutions of the problem with constant $\omega$  (Eq.~\ref{eq:WKBsol1}). Thus we guess a solution of the following form:
\begin{equation} \label{eq:WKBex2_01tris}
x(t) = A(t) \exp\left[\pm i\int_{t_0}^t \omega(s) \di s \right] \,,
\end{equation} 
where the amplitude $A(t)$ is ``slowly varying''. Note that as the argument of the exponential in \eqref{eq:WKBex2_01tris} we have the integral $\int   \omega(s) \di s$, and not the product $\omega(t) t$. Intuitively, we can think of $\int \omega(s) \di s$ as the phase of the oscillation, i.e. a number that quantifies how many oscillations occurred since the beginning of the motion. In the case $\omega=\rm constant$, the integral reduces to $\omega t$ and we recover the harmonic oscillator (Eq.~\ref{eq:WKBsol1}).

Calculating the derivatives of \eqref{eq:WKBex2_01tris} we get:
\begin{equation}
\label{eq:ddotx}
\ddot{x}(t) = \left( \ddot{A} \pm  2 i\omega \dot{A} \pm i \dot{\omega}  A - \omega^2 A \right) \exp\left[\pm i \int_{t_0}^t \omega(s) \di s \right] \,.
\end{equation}
Substituting Eqs.~\eqref{eq:WKBex2_01tris} and \eqref{eq:ddotx} into Eq.~\eqref{eq:WKBex2_01}, we obtain:
\begin{equation} \label{eq:WKBex2_02}
\ddot{A} \pm 2 i\omega \dot{A} \pm i\dot{\omega} A = 0 \,.
\end{equation}
Up to this point everything has been exact. Eq.~\eqref{eq:WKBex2_02} is completely equivalent to \eqref{eq:WKBex2_01}, and Eq.~\eqref{eq:WKBex2_01tris} can be simply viewed as a change of variable in which we replace $x$ with $A$. Now comes the WKB approximation. The essence of this approximation is that every time you take a derivative of $A(t)$ or $\omega(t)$, you get something smaller by a factor $\epsilon$, where the latter is of order of the small parameters written in Eq.~\eqref{eq:WKBcond}. In other words, we estimate the magnitudes of time derivatives by replacing $\di/\di t \sim \epsilon \omega$. Thus for example $\dot{A} \sim \epsilon \omega A $, $\dot{\omega} \sim \epsilon \omega^2 $. For the second derivatives $\ddot{A} \sim \epsilon \omega \dot{A} \sim \epsilon^2 \omega^2 A$. Using these relations, we find that the term $ \ddot{A}$ in Eq.~\eqref{eq:WKBex2_02} can be neglected compared to the others. Then Eq.~\eqref{eq:WKBex2_02} becomes:
\begin{equation} \label{eq:WKBex2_010}
\pm 2 i\omega \dot{A} \pm i\dot{\omega} A = 0 \;.
\end{equation}
This equation can be integrated and the solution is
\begin{align} \label{eq:WKBA}
A(t) = \frac{C}{\sqrt{\omega(t)}} \,.
\end{align}
Plugging \eqref{eq:WKBA} into \eqref{eq:WKBex2_01tris} we find that the general solution of Eq.~\eqref{eq:WKBex2_01} in the WKB approximation is:
\begin{equation} \label{eq:WKBgensol}
 x(t)  = \frac{C_1}{\sqrt{\omega(t)}} \exp \left[i\int_{t_0}^t \omega(s) \di s \right] +  \frac{C_2}{\sqrt{\omega(t)}} \exp \left[-i \int_{t_0}^t \omega(s) \di s \right] 
 \;,
 \end{equation}
where $C_1$ and $C_2$ are arbitrary constants. We can also write the following approximate expression for the derivative by neglecting the small terms $\dot{A}$ and $\dot{B}$:
\begin{equation} \label{eq:WKBsol2}
 \dot{x}(t) =i C_1 \sqrt{\omega(t)} \exp\left[ i \int_{t_0}^t \omega(s) \di s \right] - i C_2 \sqrt{\omega(t)} \exp\left[-i \int_{t_0}^t \omega(s) \di s \right]
\end{equation}
Note that the total energy of a simple harmonic oscillator\begin{equation}
E = \frac{1}{2} m \dot{x}^2 + \frac{1}{2} m \omega^2 x^2 \,.
\end{equation}
is not in general conserved when $\omega(t)$ is not constant. However, if we calculate $E$ using the approximate solution \eqref{eq:WKBgensol} and \eqref{eq:WKBsol2}, we find that the following quantity is constant:
\begin{equation} \label{eq:Jadiabatic}
J = \frac{E}{\omega}\,.
\end{equation}
This means that the amplitude of the oscillation becomes a function of $\omega$. If we increase $\omega$ slowly then we slowly decrease it to its original value, at the end of the process the amplitude will be the same as it was at the start. It is easy to see that this is violated if $\omega(t)$ does not change slowly (think for example of abruptly changing $\omega$ when the system passes through $x=0$: in this case the energy does not change, but $\omega$ does). The quantity $J$ is an example of an adiabatic invariant (see for example \citealt{Arnold1978} and \citealt{Landau1969} for more on adiabatic invariants).

\section{Group velocity} \label{appendix:vg}

\begin{figure}
	\includegraphics[width=\columnwidth]{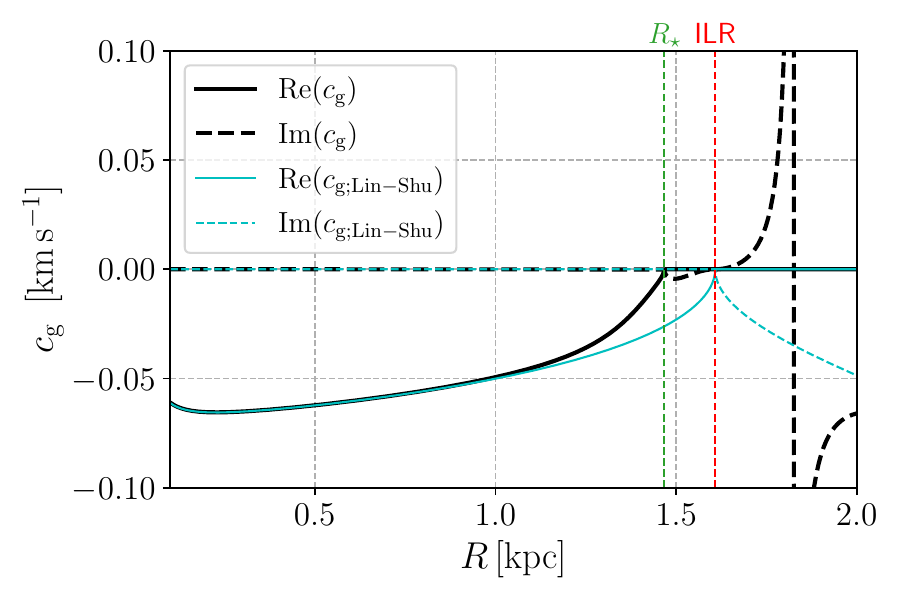}
    \caption{Group velocity of WKB waves for $\cs=10\kms$ and constant unperturbed density $\rho_0(R)=1$ (see Eq.~\ref{eq:cg}). The cyan line shows the group velocity according to the Lin-Shu dispersion relation \eqref{eq:cgLS}. The ILR and $R_\star$ are marked by vertical dashed lines.}
    \label{fig:cg}
\end{figure}

In this Appendix, we calculate the group velocity of the WKB waves. Consider a WKB solution \eqref{eq:WKB1} with $C_1\neq0$ and $C_2=0$. This is of the form:
\begin{equation} \label{eq:D01}
    g_1(R) = a(R) \exp\left[ i \phi(R) \right] \,,
\end{equation}
where
\begin{align}
    a(R) & = \frac{C_1}{\sqrt{K(R)}}, \\
    \phi(R) & =  \int_{R_0}^R  K(s) \di s \,.
\end{align}
Equation~\eqref{eq:D01} is of the same form of Equation~(1) of \cite{Toomre1969} or (1.26) of \cite{Whitham1974}. The analysis in these references shows that the group velocity, i.e. the velocity at which a wave packet travels, can be defined by isolating $\omega$ from the dispersion relation \eqref{eq:dispersionrelation} and then taking the derivative with respect to $K$:
\begin{equation} \label{eq:cg}
    \cg = \frac{\pa \omega}{\pa K} \,.
\end{equation}
Figure~\ref{fig:cg} shows the group velocity for the case $\cs=10\kms$. The group velocity of solutions with $C_1\neq 0$ and $C_2=0$ is negative, meaning that these waves travel inward, while waves with $C_1\neq 0$ and $C_2=0$ travel outwards. The group velocity of the two types of waves has the same magnitude but different sign. The group velocity loses meaning and becomes imaginary at $R>R_\star$, when the medium becomes absorbing.

The cyan line compares our group velocity with that obtained from the Lin-Shu dispersion relation, which is given by (see Equation~20 of \citealt{Goldreich1979}):
\begin{equation} \label{eq:cgLS}
c_{\rm g; Lin \mhyphen Shu} = - \frac{K_{\rm Lin\mhyphen Shu} \cs^2}{m \left(\Omega - \Omegap\right)} \,,
\end{equation}
where $K_{\rm Lin-Shu}$ is given by Eq.~\eqref{eq:KLS}. The two group velocities are similar away from $R_\star$.

\section{Angular momentum transport} \label{appendix:angularmomentum}

An equation for the angular momentum transport in a fluid disc can be obtained from Eq.~\eqref{eq:euler}. Multiplying the azimuthal component of this equation by $R$, using standard cylindrical coordinates $(R,\theta,z)$ and rearranging gives:
\begin{equation} \label{eq:angularmomentum}
\frac{\pa ( l_z )}{\pa t} + \nabla \cdot \mathbf{F}_J  = - \rho \frac{\partial \Phi}{\partial \theta} \,,
\end{equation}
where
\begin{align}
& l_z = \rho R v_\theta \,,\\ 
& \mathbf{F}_J = R \left( \rho v_\theta \bfv + P \hatephi \right)\,.
\end{align}
The quantity $l_z$ is the angular momentum per unit volume, while $\mathbf{F}_J$ is the flux of angular momentum, which is the sum of contributions due to bulk motions of the gas and pressure forces. The term $\rho \pa \Phi/\pa\theta$ is a source term representing the changes in angular momentum due to torques from the external potential. When $\pa \Phi/\pa\theta=0$, the total angular momentum of the system is conserved. Indeed, the only agent that can change the total angular momentum in our problem is the external bar potential. 

Integrating Eq.~\eqref{eq:angularmomentum} over the volume $V$ of a cylinder of radius $R_0$ and using the divergence theorem,\footnote{The divergence theorem states that for any vector-valued function $\mathbf{F}(\bfx)$:
\begin{equation}  \label{eq:divergencetheorem} 
\nonumber
\int_V \di V\, \nabla \cdot \mathbf{F} = \oint_S \di \mathbf{S} \cdot \mathbf{F}(\bfx)\,. \end{equation}} we obtain the following equation for the rate of change of the total angular momentum contained within the cylinder:
\begin{equation} \label{eq:Lzdot}
    \frac{\pa L_z}{\pa t} = - F_{\rm A} - F_{\Phi} \,,
\end{equation}
where
\begin{equation} \label{eq:Lzdef}
    L_z = \int_V \rho R v_\theta \, \di V \,,
\end{equation}
is the total $z$ angular momentum contained inside the cylinder, and
\begin{align}
F_{\rm A} & =  R^2 \int_{-\infty}^{\infty} \di z \int_{0}^{2\pi} \di \theta \rho v_\theta v_R  \label{eq:FA} \\
F_{\Phi} & = \int_V \rho \frac{\pa \Phi}{\pa \theta}\, \di V\,, \label{eq:FG}
\end{align}
are the fluxes of angular momentum in and out of the cylinder.

Eq.~\eqref{eq:Lzdot} states that the change in the total angular momentum of the gas contained within the cylinder is the sum of two contributions: $F_A$, the angular momentum flux due to advection, and $F_{\rm \Phi}$, the gravitational torques from the external bar potential.

The quantity $F_A$ appears in \cite{Goldreich1979} as their Eq.~(26). $F_{\rm A}>0$ means that material inside the cylinder is losing angular momentum.  Notice that even in a steady-state, in which single fluid elements neither gain nor lose angular momentum on average, it is nevertheless possible that $F_{\rm A} \neq 0$. This can happen if fluid elements carry more angular momentum on their outward journey (as they are exiting the cylinder) than on their return. This type of transport has been named \emph{lorry transport} by \cite{Lynden-Bell1972}, who explained how fluid elements can ``transport angular momentum just as a system of lorries can transport coal without accumulating a growing store on the lorries themselves''. This is similar to a plane sound wave transporting linear momentum in a steady-state situation.

In perturbed 2D discs (where quantities are expanded as in Eq.~\ref{eq:rho1}) Eq.~\eqref{eq:FA} can be simplified to (see Eq.~J.16 in \citealt{BT2008}):
\begin{equation} \label{eq:FA2}
F_{\rm A} =  R^2 \rho_0 \int_{0}^{2\pi} \di \theta v_{\theta1} v_{R1} \,.
\end{equation}
Using Eqs.~\eqref{eq:vR1} and \eqref{eq:vtheta1}, we can rewrite this as (see Eq.~J.19 in \citealt{BT2008}):
\begin{equation} \label{eq:FA3}
F_{\rm A} =  \operatorname{Re}\left\{ \frac{\pi R \rho_0 i m}{D} \left[ \left(\Phi_1 + h_1\right)^* \frac{\di}{\di R}\left(\Phi_1 + h_1 \right) \right] \right\}\,.
\end{equation}
To evaluate the flux of angular momentum associated with the WKB waves (Eq.~\ref{eq:WKB1}), we substitute Eq.~\eqref{eq:WKB1} into Eq.~\eqref{eq:FA3} using Eq.~\eqref{eq:g1} and set $\Phi_1=0$. After a straightforward calculation we obtain:
\begin{equation}
\label{eq:FAWKB}
F_{\rm A} = m\pi \left(\left|C_2\right|^2 - \left|C_1\right|^2 \right) {\rm sgn}\left(D\right) \;.
\end{equation}
At the radii of interest (i.e., inside the ILR), one has ${\rm sgn}(D)=-1$.

\section{Details on the calculations of waves excited at a sharp edge}
\label{appendix:calculations}

\subsection{Derivation of Eqs.~\eqref{eq:A1}-\eqref{eq:Cinout}} \label{appendix:calc1}

In a neighbourhood of the point $R=R_{\rm out}$ we can approximate Eq.~\eqref{eq:g1neargen} as
\begin{align}
g_1(R) & = e^{iKR} \left[A_1-\frac{Q_{\rm out}}{2K^2}e^{-iKR_{\rm out}} - \frac{i}{2K}\int_{R_0}^{R_{\rm out}}Q(s)e^{-iKs}\di s \right] + \nonumber\\
& + e^{-iKR} \left[A_2-\frac{Q_{\rm out}}{2K^2}e^{iKR_{\rm out}} + \frac{i}{2K}\int_{R_0}^{R_{\rm out}}Q(s)e^{iKs}\di s \right] + \nonumber\\
\label{eq:g1out}
& + \frac{Q_{\rm out}}{K^2} \qquad \text{(neighbourhood of }R_{\rm out}\rm) \;,
\end{align}
where $Q_{\rm out}=Q(R_{\rm out})$. Since the waves are travelling outwards at $R=R_{\rm out}$, the term proportional to $e^{iKR}$ should vanish. This condition gives Eq.~\eqref{eq:A1}.

Similarly, in a neighbourhood of the point $R=R_{\rm in}$ we can approximate Eq.~\eqref{eq:g1neargen} as
\begin{align}
g_1(R) & = e^{iKR} \left[A_1-\frac{Q_{\rm in}}{2K^2}e^{-iKR_{\rm in}} - \frac{i}{2K}\int_{R_0}^{R_{\rm in}}Q(s)e^{-iKs}\di s \right] + \nonumber\\
& + e^{-iKR} \left[A_2-\frac{Q_{\rm in}}{2K^2}e^{iKR_{\rm in}} + \frac{i}{2K}\int_{R_0}^{R_{\rm in}}Q(s)e^{iKs}\di s \right] + \nonumber\\
\label{eq:g1in}
& + \frac{Q_{\rm in}}{K^2} \qquad \text{(neighbourhood of }R_{\rm in}\rm) \;,
\end{align}
where $Q_{\rm in}=Q(R_{\rm in})$. Since the waves are travelling inwards at $R=R_{\rm in}$, the term proportional to $e^{-iKR}$ should vanish. This condition gives Eq.~\eqref{eq:A2}.

Substituting Eqs.~\eqref{eq:A1} and \eqref{eq:A2} into Eq.~\eqref{eq:g1out} and Eq.~\eqref{eq:g1in} respectively we find
\begin{align}
g_1(R) & = e^{-iKR} \left[ \frac{i}{2K}\int_{R_{\rm in}}^{R_{\rm out}}Q(s)e^{iKs}\di s \right] + \nonumber\\
& + e^{-iKR} \left[\frac{Q_{\rm in}}{2K^2}e^{iKR_{\rm in}}-\frac{Q_{\rm out}}{2K^2}e^{iKR_{\rm out}} \right] + \nonumber\\
\label{eq:g1outfinal}
& + \frac{Q_{\rm out}}{K^2} \qquad \text{(neighbourhood of }R_{\rm out}\rm) \;,
\end{align}
and
\begin{align}
g_1(R) & = e^{iKR} \left[ \frac{i}{2K}\int_{R_{\rm in}}^{R_{\rm out}}Q(s)e^{-iKs}\di s \right] + \nonumber\\
& +e^{iKR} \left[\frac{Q_{\rm out}}{2K^2}e^{-iKR_{\rm out}}-\frac{Q_{\rm in}}{2K^2}e^{-iKR_{\rm in}} \right] +  \nonumber\\
\label{eq:g1infinal}
& + \frac{Q_{\rm in}}{K^2} \qquad \text{(neighbourhood of }R_{\rm in}\rm) \;.
\end{align}
Matching Eqs.~\eqref{eq:g1outfinal} and \eqref{eq:g1infinal} with Eq.~\eqref{eq:g1WKBinout}, one obtains Eq.~\eqref{eq:Cinout}.

\subsection{Derivation of Eq.~\eqref{eq:Capprox}}
\label{sec:impulse}

We approximate Eq.~\eqref{eq:Cinout} as follows. First, we neglect the terms proportional to $Q_{\rm in}$ and $Q_{\rm out}$ because $Q$ varies rapidly at radii $R_{\rm in}<R<R_{\rm out}$. We obtain
\begin{equation}
\left|C_{\rm in}\right|\simeq\left|C_{\rm out}\right|\simeq\frac{1}{2K^{1/2}} \left|\int_{R_{\rm in}}^{R_{\rm out}}Q(s)e^{-iKs}\di s \right| \;.
\end{equation}
Second, the exponential $\exp(- iKs)$ is nearly constant as we have assumed $|R_{\rm out}-R_{\rm in}|\sim \lambda \sim 1/K$, so we can write
\begin{equation}
\label{eq:Capproxapp}
\left|C_{\rm in}\right|\simeq\left|C_{\rm out}\right|\simeq\frac{1}{2K^{1/2}} \left|\int_{R_{\rm in}}^{R_{\rm out}}Q(s)\di s \right| \;.
\end{equation}
We have
\begin{equation}
\label{eq:Qint}
\int_{R_{\rm in}}^{R_{\rm out}} Q(s)\di s = I_1+I_2+I_3+I_4\;,
\end{equation}
where
\begin{align}
I_1 & = -\int_{R_{\rm in}}^{R_{\rm out}} \di s\left(\frac{s\rho_0}{\left|D\right|}\right)^{1/2}\frac{\di^2\Phi_1}{\di s^2} \\
I_2 & = -\int_{R_{\rm in}}^{R_{\rm out}} \di s \left(\frac{s\rho_0}{\left|D\right|}\right)^{1/2} \frac{\di}{\di s}\left[\log\left(\frac{s\rho_0}{D}\right)\right]\frac{\di\Phi_1}{\di s} \\
I_3 & = -\int_{R_{\rm in}}^{R_{\rm out}} \di s \left(\frac{s\rho_0}{\left|D\right|}\right)^{1/2} \frac{2\Omega}{s\left(\Omega-\Omega_{\rm p}\right)} \frac{\di}{\di s}\left[\log\left(\frac{\rho_0\Omega}{D}\right)\right] \Phi_1 \\
I_4 & = \int_{R_{\rm in}}^{R_{\rm out}} \di s \left(\frac{s\rho_0}{\left|D\right|}\right)^{1/2} \frac{m^2\Phi_1}{s^2} \;.
\end{align}
Since $|R_{\rm out}/R_{\rm in}-1|\ll 1$ and far from the ILR the integrand is bounded, we have $I_1\simeq I_4\simeq 0$.

We calculate $I_2$ and $I_3$ below. The idea is to integrate by parts in order to isolate the integral of a bounded function. We have
\begin{align}
I_2 & = -2\int_{R_{\rm in}}^{R_{\rm out}}\di s \frac{\di}{\di s}\left(\frac{s\rho_0}{\left|D\right|}\right)^{1/2}\frac{\di\Phi_1}{\di s}=\nonumber\\
& = -2\left[ \left(\frac{R\rho_0}{\left|D\right|}\right)^{1/2}\frac{\di\Phi_1}{\di R}\right]_{R=R_{\rm out}} +2\left[ \left(\frac{R\rho_0}{\left|D\right|}\right)^{1/2}\frac{\di\Phi_1}{\di R}\right]_{R=R_{\rm in}}+\nonumber\\
& + 2\int_{R_{\rm in}}^{R_{\rm out}}\di s \left(\frac{s\rho_0}{\left|D\right|}\right)^{1/2}\frac{\di^2\Phi_1}{\di s^2} = \nonumber\\
\label{eq:I2}
& = 2\left[ \left(\frac{R\rho_0}{\left|D\right|}\right)^{1/2}\frac{\di\Phi_1}{\di R}\right]_{R=R_{\rm in}} \;,
\end{align}
and
\begin{align}
I_3 & = -4\int_{R_{\rm in}}^{R_{\rm out}} \di s \left[\frac{\Omega^{1/2}\Phi_1}{s^{1/2}\left(\Omega-\Omega_{\rm p}\right)}\right] \frac{\di}{\di s}\left(\frac{\rho_0\Omega}{\left|D\right|}\right)^{1/2} =\nonumber\\
& = - \left[\frac{4\Omega}{\Omega-\Omega_{\rm p}}\left(\frac{R\rho_0}{\left|D\right|}\right)^{1/2}\frac{\Phi_1}{R}\right]_{R=R_{\rm out}} + \nonumber\\
& + \left[\frac{4\Omega}{\Omega-\Omega_{\rm p}}\left(\frac{R\rho_0}{\left|D\right|}\right)^{1/2}\frac{\Phi_1}{R}\right]_{R=R_{\rm in}} +\nonumber\\
& + 4\int_{R_{\rm in}}^{R_{\rm out}} \di s \left(\frac{\rho_0\Omega}{\left|D\right|}\right)^{1/2} \frac{\di}{\di s} \left[\frac{\Omega^{1/2}\Phi_1}{s^{1/2}\left(\Omega-\Omega_{\rm p}\right)}\right]  =\nonumber\\
\label{eq:I3}
& = \left[\frac{4\Omega}{\Omega-\Omega_{\rm p}}\left(\frac{R\rho_0}{\left|D\right|}\right)^{1/2}\frac{\Phi_1}{R}\right]_{R=R_{\rm in}} \;,
\end{align}
where we have used the fact that $\rho_0(R_{\rm in})\gg\rho_0(R_{\rm out})$. Substituting Eqs.~\eqref{eq:I2} and \eqref{eq:I3} into Eq.~\eqref{eq:Qint}, we find
\begin{equation}
\label{eq:Qfinal}
\int_{R_{\rm in}}^{R_{\rm out}} Q(s)\di s = 2\left[\left(\frac{R\rho_0}{\left|D\right|}\right)^{1/2}\left(\frac{\di\Phi_1}{\di R} +\frac{2\Omega}{\Omega-\Omega_{\rm p}}\frac{\Phi_1}{R}\right)\right]_{R=R_{\rm in}}\;.
\end{equation}
Substituting Eq.~\eqref{eq:Qfinal} into Eq.~\eqref{eq:Capproxapp}, we obtain Eq.~\eqref{eq:Capprox}.

\bsp	
\label{lastpage}
\end{document}